\documentclass[aps,prb,reprint,superscriptaddress,showpacs]{revtex4-2}

\usepackage{amsmath}
\usepackage{xcolor}
\usepackage{graphicx}
\usepackage{dcolumn}
\usepackage{bm}
\usepackage{hyperref}
\usepackage[mathlines]{lineno}
\usepackage{float}
\usepackage{comment}

\begin{document}


\title{Filtering Spin and Orbital Moment 
in Centrosymmetric Systems}

\author{Luciano Jacopo D'Onofrio}
\affiliation{CNR-SPIN, c/o Universit\`a di Salerno, IT-84084 Fisciano (SA), Italy}

\author{Maria Teresa Mercaldo}
\affiliation{Dipartimento di Fisica ``E. R. Caianiello", Universit\`a di Salerno, IT-84084 Fisciano (SA), Italy}

\author{Wojciech Brzezicki}
\affiliation{Institute of Theoretical Physics, Jagiellonian University, ulic, S. \L{}ojasiewicza 11, PL-30348 Krak\'ow, Poland}
\affiliation{International Research Centre MagTop, Institute of Physics, Polish Academy of Sciences, Aleja Lotnik\'ow 32/46, PL-02668 Warsaw, Poland}

\author{Adam Kłosiński}
\affiliation{International Research Centre MagTop, Institute of Physics, Polish Academy of Sciences, Aleja Lotnik\'ow 32/46, PL-02668 Warsaw, Poland}

\author{Federico Mazzola}
\affiliation{CNR-SPIN, c/o Complesso di Monte S. Angelo, IT-80126 Napoli, Italy}

\author{Carmine Ortix}
\affiliation{Dipartimento di Fisica ``E. R. Caianiello", Universit\`a di Salerno, IT-84084 Fisciano (SA), Italy}

\author{Mario Cuoco}
\affiliation{CNR-SPIN, c/o Universit\`a di Salerno, IT-84084 Fisciano (SA), Italy}

\begin{abstract}
The control of spin and orbital angular momentum without relying on magnetic materials
is commonly accomplished by breaking of inversion symmetry, which enables charge-to-spin conversion and 
spin selectivity in electron transfer processes occurring in chiral media.
In contrast to this perspective, 
we show that orbital moment filtering can be accomplished in centrosymmetric systems: the electron states can be selectively manipulated allowing for the preferential transfer of electrons with a particular orbital momentum orientation. We find that orbital moment filtering is indeed efficiently controlled through orbital couplings that break both mirror and rotational symmetries.
We provide the symmetry conditions required for the electron transmission to achieve orbital filtering and relate them to the orientation of the orbital moment.
The presence of atomic spin-orbit interaction in the centrosymmetric transmission medium leads to the selective filtering of spin and orbital moments. 
Our findings allow to identify optimal regimes for having highly efficient simultaneous spin and orbital moment filtering.


\end{abstract}

\maketitle



\section{Introduction}
A key factor in the progress of spintronics and spin-orbitronics, as well as in the development of new functionalities in solid materials, is the capacity to encode, manipulate, and store information using spin and orbital angular momentum. 
\textcolor{black}{Apart from the spin Hall effect that can generally lead to transverse spin accumulation of a sample carrying electric current \cite{Dyakonov1971,Hirsch1999,Valenzuela2006},}
\textcolor{black}{a promising and widely utilized approach for designing spin functionalities 
and facilitating charge-to-spin conversion involves strong spin-orbit coupling and acentric materials.}
The spin-momentum locking \cite{Rashba1960,Dresselhaus1955}, due to the breaking of inversion symmetry, allows for the control of the spin angular moment by transforming, for instance, an unpolarized charge current into a spin density through the Edelstein effect \cite{Sanchez2013,Varotto2022,soumyanarayanan_nat16}. 
It has recently been discovered that many spin-related phenomena can be traced back to orbital mechanisms \cite{Jo2024} even without the presence of atomic spin-orbit coupling. In this context, 
orbital-based couplings can give rise to orbital Hall \cite{Tanaka2008,Kontani2009,Go2018,Cysne2022,Sala2023,Lyalin2023,Choi2023}, orbital Edelstein \cite{Yoda_2018,Johansson2021, Go2017, Salemi2019, Chirolli2022}, inverse orbital Edelstein effects \cite{ElHamdi2023}, orbital photocurrents \cite{Adamantopoulos2024}, and orbital sources of Berry curvature \cite{les23,Mercaldo2023}.

In systems with weak spin-orbit coupling when inversion symmetry breaking occurs alongside the absence of other crystalline symmetries, as seen in chiral structures, other effects can emerge and electron transmission demonstrates selective filtering based on the spin of the electrons. This effect, referred to as chiral induced spin selectivity (CISS), arises from the polarization of electron spins by chiral molecules \cite{Ray1999-ex,Naaman2019}. Although CISS has primarily been observed in chiral molecules, studies have revealed that chiral crystals can also generate spin-polarized states through the injection of charge currents \cite{Furukawa2017,Calavalle2022,Yang2021}. Despite several investigations there is no consensus on the mechanisms behind the CISS. However, orbital texture and orbital polarization effects have been proposed to induce spin-polarization both in chiral systems but also in achiral materials that, however, lack a center of inversion \cite{Liu2021}. 

All of these observations indicate that breaking inversion symmetry, along with the presence of spin and orbital textures in momentum space, plays a crucial role for controlling electron spin and orbital moments. \textcolor{black}{Nevertheless, despite not being directly related to spin and orbital filtering effects, there are certain cases where orbital dynamics \cite{Han2022} and hidden spin or orbital polarization can arise in centrosymmetric systems \cite{Zhang2014,Riley2014,cappelluti2024}.}
\textcolor{black}{
Notably, spin-filtering has been successfully accomplished using centrosymmetric molecules \cite{Pal2019}, demonstrating that a chiral structure is not a requisite for the transmission of spin-polarized electrons.}
Therefore, uncovering the mechanisms that enable the manipulation and control of spin and orbital moments from a symmetry perspective is essential.

\begin{figure}
\centering
\includegraphics[width=1.\columnwidth]{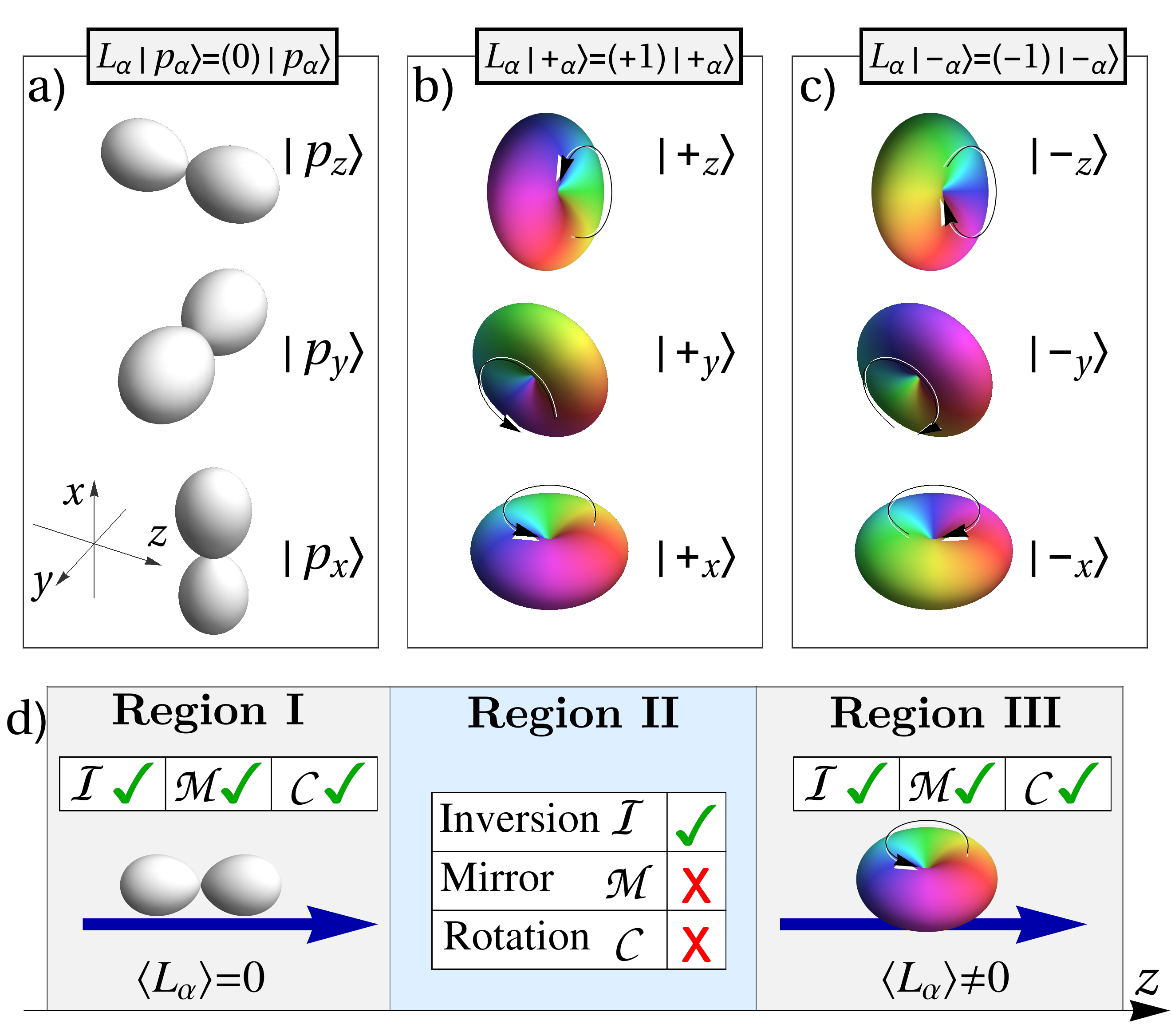}
\caption{Schematic of the orbital configurations involved in the transmission process along a given direction ($z$). Illustration of 
atomic orbitals $\{|p_z\rangle,|p_y\rangle,|p_x\rangle \}$ having zero (a) and nonvanishing (b),(c) projections of the orbital moment along the $x,y,z$ symmetry directions. Sketch of (b) $+1$ and (c) $-1$ orbitally polarized configurations.
Clockwise or anticlockwise arrows are used to label states with opposite orbital moment along a given direction. 
$|\pm_z \rangle=\frac{1}{\sqrt{2}}(\mp i |p_x\rangle +|p_y\rangle)$ and similarly for the other components. (d) 
Illustration of electron transmission: 
An electron is initialized 
in a state with zero orbital moment in $\text{I}$ with preserved inversion ($\mathcal{I}$), mirror ($\mathcal{M}$) and rotational ($\mathcal{C}$) symmetries. It is then transmitted through region $\text{II}$, where $\mathcal{M}$ and $\mathcal{C}$ are broken \textcolor{black}{or the mirror/rotational symmetries are not equivalent to those in the region $\text{I}$}, before entering region $\text{III}$ that has the same symmetry properties of $\text{I}$. The resultant state in $\text{III}$ exhibits a non-zero orbital moment.}
\label{fig:1}
\end{figure}

In this paper, we address this fundamental issue by exploring how spin and orbital moment filtering can take place in centrosymmetric systems.
By utilizing orbital couplings that preserve inversion symmetry and break both mirror and rotational symmetries, we demonstrate the selective transfer of electrons with a specific orientation of orbital momentum. Our findings reveal the connection between symmetry-breaking interactions and the manipulation of the orbital moment orientation for the transmitted electronic states. Consequently, spin filtering can be realized by introducing a non-zero atomic spin-orbit coupling (SOC) in the centrosymmetric transmission medium. Our analysis indicates the path to achieve high-efficiency filtering in the generation of both spin- and orbitally polarized currents
by injecting unpolarized electrons into a centrosymmetric medium that \textcolor{black}{exhibits distinct mirror and rotational symmetry properties with respect to the injection region (Fig. \ref{fig:1})}.

\textcolor{black}{This paper is organized as follows. In Section \ref{sec:II}, we introduce a model for electron propagation through a centrosymmetric, time-reversal invariant medium, and define a multi-orbital Hamiltonian describing the effects of symmetry-breaking orbital couplings. We discuss the symmetry properties of the system and outline the method implemented to solve the transmission problem and evaluate the resulting orbital and spin polarization. Section \ref{sec:III} presents the main results, including conditions for orbital and spin filtering, the role of symmetry-breaking couplings, and their interplay with atomic spin-orbit interaction. The response is analyzed as a function of orbital configuration, system parameters, and symmetry constraints. Finally, we summarize our conclusions in Section \ref{sec:IV}.}

\begin{figure*}
\centering
\includegraphics[width=1\textwidth]{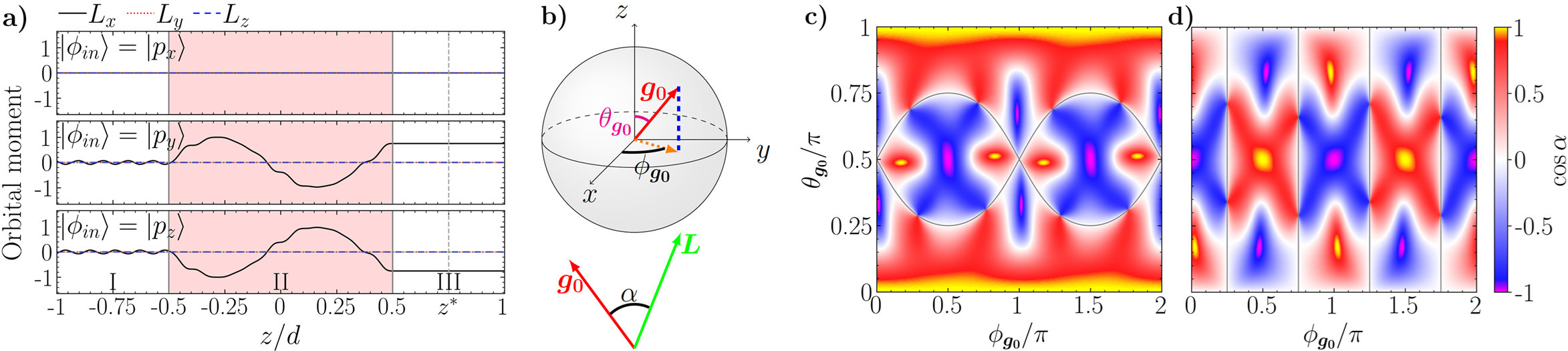}
\caption{
(a) Spatial profile of the orbital moment associated to the resulting state of the electron transmission problem in the regions $\text{I}$, $\text{II}$ (pink shaded area), $\text{III}$, considering different $p$-orbital configurations as input configurations ($|\phi_{in}\rangle$)\textcolor{black}{, $g_{0x}/\epsilon_0=0.5$, and $d/d_0=20$, with $\epsilon_0=38$ meV and $d_0=1$ nm. Here, $l_F\sim d/4$, where $l_F$ is the average Fermi wavelength in the region $\text{II}$}.
(b) Illustration of the spherical parametrization, through the variables $(\theta_{\boldsymbol{g_0}},\phi_{\boldsymbol{g_0}})$, of the $\boldsymbol{g_0}$-vector coupling in the region $\text{II}$ that breaks mirror and rotational symmetries while keeping the inversion symmetry. The angle $\alpha$ is introduced to evaluate the orientation of the orbital moment filtered in the region $\text{III}$ with respect to the character of the $\boldsymbol{g_0}$ coupling \textcolor{black}{assuming a representative amplitude $|\boldsymbol{g_0}|/\epsilon_0=0.5$ and $d/d_0=20$}.
The evaluation of $\boldsymbol{L}$ for the outgoing state refers to a given position $z^*$ in the region $\text{III}$ as indicated in the panel (a).
Contour maps of $\cos\alpha$ for an input state with $p_x$ (c) and $p_z$ (d) type. Solid lines indicate the values of $\boldsymbol{g_0}$ coupling for which $\boldsymbol{g_0}$ is perpendicular to $\boldsymbol{L}$ \textcolor{black}{as dictated by the corresponding symmetries in orbital space; other points for which $\boldsymbol{g_0}\perp \boldsymbol{L}$ holds are accidental, due to specifics of the corresponding scattering processes.} The results indicate that $\boldsymbol{g_0}$ and $\boldsymbol{L}$ are never collinear except for distinct points: \textcolor{black}{some are symmetry-induced, corresponding to the single-component $\boldsymbol{g_0}$, while others depend on the details of the scattering problem under study.}
}
\label{fig:2}
\end{figure*}

\section{Model and methodology}\label{sec:II}
To illustrate how electron transmission in a centrosymmetric medium that is time-reversal preserving can selectively filter spin and orbital moments, we examine an $L=1$ multi-orbital electronic system and incorporate orbital couplings to address the breaking of both mirror and rotational symmetries. The electronic states are expressed in terms of atomic orbitals $\{|p_z\rangle,|p_y\rangle,|p_x\rangle \}$ \textcolor{black}{oriented along the axes of given reference system $Oxyz$ (Fig. \ref{fig:1}a). Here, we introduce the following representation \begin{equation}
\left|p_{z}\right\rangle=\left(\begin{array}{c}
1\\
0\\
0
\end{array}\right),
\left|p_{y}\right\rangle=\left(\begin{array}{c}
0\\
1\\
0
\end{array}\right),
\left|p_{x}\right\rangle=\left(\begin{array}{c}
0\\
0\\
1
\end{array}\right).
\end{equation}
In the manifold spanned by the $p$-orbitals the components of the atomic
orbital moment $\hat{\boldsymbol{L}}$ can be written as
\begin{align}
\hat{L}_{x} & =\left(\begin{array}{ccc}
0 & -i & 0\\
i & 0 & 0\\
0 & 0 & 0
\end{array}\right),\thinspace\hat{L}_{y}=\left(\begin{array}{ccc}
0 & 0 & -i\\
0 & 0 & 0\\
i & 0 & 0
\end{array}\right),\nonumber \\
 & \hat{L}_{z}=\left(\begin{array}{ccc}
0 & 0 & 0\\
0 & 0 & -i\\
0 & i & 0
\end{array}\right),
\end{align}
\noindent where the usual commutation rule $\left[\hat{L}_{i},\hat{L}_{j}\right]=i\varepsilon_{ijk}\hat{L}_{k}$
is satisfied for any triplet $i,j,k=x,y,z$, with $\hat{L}_{i}\left|p_{i}\right\rangle=0$.}
For the spin angular momentum, we use the standard representation with  $\hat{S}_{i}=\frac{1}{2} \hat{\sigma}_{i}$, and $\hat{\sigma}_{i}$ being the Pauli matrices.
The orbital configurations of the electronic states can be also characterized in terms of the products of orbital moment. 
Given that the orbital angular momentum has pseudovector characteristics with magnetic dipole properties, their nontrivial combination can be classified as an orbital quadrupole. For convenience and clarity we consider a transmission system featuring a single propagation direction ($z$) and comprises three regions, each exhibiting distinct symmetry characteristics (Fig. \ref{fig:1}d). The outer regions ($\text{I}$ and $\text{III}$) preserve inversion, mirror and twofold rotational symmetries. Instead, in the region $\text{II}$ we introduce orbital couplings that break only mirror and rotation while keeping inversion symmetry. 
Hence, the Hamiltonian can be generally expressed as $\hat{\mathcal{H}}=\hat{\mathcal{H}}\left(z\right)$:
\begin{equation}
\hat{\mathcal{H}}\left(z\right)=\begin{cases}
\hat{\mathcal{H}}^{\text{I}}\left(z\right) & z<-d/2\\
\hat{\mathcal{H}}^{\text{II}}\left(z\right) & -d/2\leq z\leq d/2\\
\hat{\mathcal{H}}^{\text{III}}\left(z\right) & z>d/2
\end{cases} \,,
\end{equation}
where $d$ is the width of the region $\text{II}$.  
\textcolor{black}{The Hamiltonian in $\text{I}$ and $\text{III}$ is given by}
\begin{align}\textcolor{black}{
\hat{\mathcal{H}}^{A}\left(z\right) =\hat{p}_{z}^{2}\left(\sum_{i=x,y,z}a_{i}^{A}\hat{L}_{i}^{2}\right) + \sum_{i=x,y,z}\Delta_{i}^{A}\hat{L}_{i}^{2},}
\end{align}
\textcolor{black}{with $A=\text{I}$ or $\text{III}$,  $\hat{p}_{z}$ is
the electronic linear momentum operator along the $z$-axis, $a_{i}^{A}$ are the orbital dependent mass and $\Delta_{i}^{A}$ the crystalline field amplitudes, respectively. 
Here, the terms proportional to $\hat{L}_{i}^{2}$ are crystal fields that preserve mirror, twofold rotation and inversion symmetries.
For the region $\text{II}$ we have that}
\begin{align}
\hat{\mathcal{H}}^{\text{II}}\left(z\right) & =\hat{p}_{z}^{2}\left(\sum_{i=x,y,z}a_{i}^{\text{II}}\hat{L}_{i}^{2}+\boldsymbol{g}\cdot\hat{\boldsymbol{\chi}}\right)+\nonumber\\ & + \sum_{i=x,y,z}\Delta_{i}^{\text{II}}\hat{L}_{i}^{2}+\boldsymbol{g_0}\cdot\hat{\boldsymbol{\chi}},
\end{align}
where $\boldsymbol{g}$, $\boldsymbol{g_0}$ are the strength of orbital couplings that can break mirror and twofold rotational symmetries with respect to the $x,y,z$ reference axes. 
These symmetry breaking interactions are expressed through the operators $\hat{\boldsymbol{\chi}}=(\hat{{\chi}}_x,\hat{{\chi}}_y,\hat{{\chi}}_z)$ with $\hat{{\chi}}_x=\hat{L}_{y} \hat{L}_{z}+\hat{L}_{z} \hat{L}_{y}$, and the other components obtained by cyclic index permutation. 
Both $\boldsymbol{g}$ and $\boldsymbol{g_0}$ can lower the symmetry while preserving the spatial inversion: $\boldsymbol{g}$ acts through a change in the electron mass, while $\boldsymbol{g_0}$ is a crystalline field potential. Since we are dealing with a one-dimensional transmission, the relevant rotation symmetry is the twofold $\hat{\mathcal{C}}_2$ transformation that rotates by $\pi$ around a given axis. Furthermore, we define the symmetry transformation $\hat{\mathcal{M}}_i$ as the mirror operation relative to the plane that is perpendicular to the $i$-axis and passing through the origin. We recall that the inversion symmetry is expressed as $\hat{\mathcal{I}}=\hat{\mathcal{M}}_i \hat{\mathcal{C}}_{2i}$. \textcolor{black}{In addition, we represent with $\hat{M}_{ij}$ ($\hat{M}_{\overline{ij}}$) the symmetry operation (along with its representation) in the orbital space that exchanges $\hat{L}_i$ with $\left(-\right)\hat{L}_j$, i.e., $\hat{M}_{ij}^{-1}\hat{L}_i\hat{M}_{ij}=\hat{L}_j$ ($\hat{M}_{\overline{ij}}^{-1}\hat{L}_i\hat{M}_{\overline{ij}}=-\hat{L}_j$).}
Considering that the components of $\hat{\boldsymbol{L}}$ transform as pseudovectors upon mirror transformation, one can immediately show that the term $\hat{{\chi}}_x$  preserves $\hat{\mathcal{I}}$ as well as $\hat{\mathcal{M}}_x$ and the twofold rotational symmetry $\hat{\mathcal{C}}_{2x}$ with respect to the $x$-axis. Similar considerations apply to
$\hat{{\chi}}_y$ and $\hat{{\chi}}_z$.

The transmission problem is then analyzed in the standard way by
solving the Schr\"{o}dinger equation in the three regions and imposing proper boundary conditions at their interfaces (see Appendix \ref{app:B}).
We assume that the initial state ($\phi_{in}$) is set up in a configuration with specific orbital occupation corresponding to the zero orbital moment $p_x$, $p_y$, and $p_z$ states. The aim is to assess whether there is a probability to transmit an electron with non zero orbital moment in the region III. 
In order to visualize the character of the transmitted electron current, it is convenient to evaluate the spin and orbital moment polarization at any given position $z$ which is expressed as $\boldsymbol{\mathcal{O}}^{}\left(z\right)=\left(\mathcal{O}_{x}^{}\left(z\right),\mathcal{O}_{y}^{}\left(z\right),\mathcal{O}_{z}^{}\left(z\right)\right)$ with $\mathcal{O}=L$, $S$ being the orbital and spin moment, respectively. 
All the outcomes are expressed in terms of a characteristic length scale $k_0$ and energy $\epsilon_{0}=\hbar^{2} k_{0}^{2}/2m$, such that $k_0 d_0 =1$, with $\hbar$ the reduced Planck constant and $m$ the electron mass, respectively. Here, we choose $d_0=1$ nm and, thus, $\epsilon_0=38$ meV. Moreover, we consider one representative case for the energy $\mu$ of the electronic state, i.e. $\mu/\epsilon_{0}=2$. A change in $\mu$ does not alter the qualitative behavior of the results.

\section{Results}\label{sec:III}


\begin{figure}[!t]
\centering
\includegraphics[width=1\linewidth]{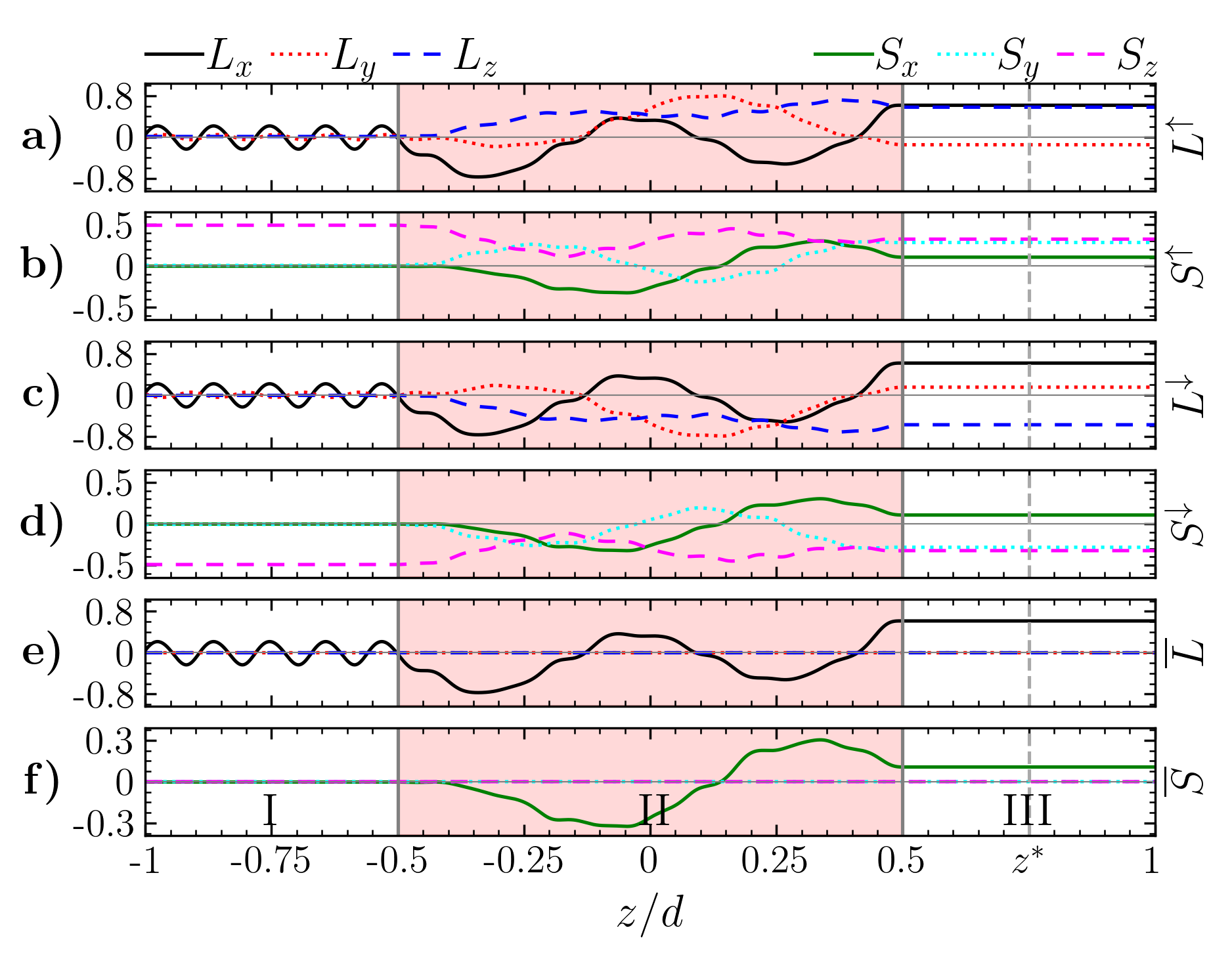}
\caption{
Spatial profile of orbital ((a) $L_\uparrow$ and (c) $L_\downarrow$) and spin ((b) $S_\uparrow$ and (d) $S_\downarrow$) moments associated to the resulting state of the electron transmission problem in the regions $\text{I}$, $\text{II}$ (pink shaded area), $\text{III}$, considering a $p_z$-orbital configuration as input ($|\phi_{in}\rangle$) and spin states with polarization oriented along the $z$-axis, i.e. $|\uparrow_{z}\rangle$ and $|\downarrow_{z}\rangle$.
\textcolor{black}{The parameters are $g_{0x}/\epsilon_{0}=0.5$, $g_{0y}=g_{0z}=0$, $l_F\sim d/4$, $\lambda/\epsilon_{0}=0.6$, and $d/d_0=20$ ($l_F$ is the mean Fermi wavelength, $\epsilon_0=38$ meV and $d_0=1$ nm).}  
In panels (e) and (f) we report the spatial profile of 
$\overline{L}=(L^{\uparrow}+L^{\downarrow})/2$ and $\overline{S}=(S^{\uparrow}+S^{\downarrow})/2$, averaging over the two spin channels considering that the electrons are prepared as an equal mixture of spin up and down configurations: a spin unpolarized charge current is injected into the region $\text{II}$ and converted into a spin and orbitally polarized current with net moment oriented along the $x$-axis. 
}
\label{fig:3}
\end{figure}

Let us now examine the electron transmission. \textcolor{black}{For convenience and clarity we present results with vanishing $\boldsymbol{g}$ as it does not alter the qualitative behavior of the outcome as explicitly shown in Appendix \ref{app:C}. Additionally, the use of $\boldsymbol{g_0}$ is convenient because it can be directly connected to material properties, as it results in orbital dependent energy splitting at the center of the Brillouin zone.} 
The Hamiltonian in $\text{I}$ and $\text{III}$ has vanishing $\boldsymbol{g_0}$ couplings and, for clarity, we assume that there is no crystal field potential. Splitting of orbital degeneracy due to $\Delta_i^{\text{I},\text{II}}$ 
, along with mass anisotropy given by $a_i^{\text{II}}$, does not affect the qualitative nature of the results (see Appendix \ref{app:C}).
We consider
a representative case where only one component \textcolor{black}{(e.g. $g_{0x}$) of the $\boldsymbol{g_0}$} couplings is nonzero and the regions $\text{I}$ and $\text{III}$ have equal electronic parameters so that inversion symmetry holds for the global system too. 
In this configuration, when $\phi_{in}$ in region $\text{I}$ exhibits only $p_x$ character, the operator $\hat{{\chi}}_x$ is ineffective. As a result, there is no orbital mixing during propagation, and the outgoing state retains a zero orbital momentum (Fig. \ref{fig:2}a).
Instead for 
$\phi_{in}$ prepared in an orbital configuration of $p_y$ or $p_z$ type, we observe that the probability of obtaining a state with orbital polarization $+_x$ differs from that of $-_x$ in the outgoing configuration. This leads to a non-zero orbital moment $L_x$ for the transmitted state (see Fig. \ref{fig:2}a). 
One can analytically determine the transmission matrix (see Appendix \ref{app:C}) and show the conversion from $p_y$ to states with nonvanishing orbital moment. 
These results are among the key outcomes of our study, as they illustrate the orbital moment being filtered through a centrosymmetric medium that maintains at least one mirror and a twofold rotational symmetry. 
We note that inversion symmetry prohibits the presence of a uniform orbital moment in response to the injected current within the system. However, as just demonstrated, it does not eliminate the possibility of having a finite spatial gradient of the orbital moment in response to a charge current. 

Now, a question arises of how the orientation of the filtered moment relates to the orbital couplings that govern the overall breaking of mirror and rotational symmetries. 
To achieve this, we set a given amplitude for the \textcolor{black}{$\boldsymbol{g_0}$ coupling and use a spherical 
representation for the 
$(g_{0x}, g_{0y}, g_{0z})$ components} (Fig. \ref{fig:2}b). Next, we determine how the orbital moment in region $\text{III}$ correlates with the orbital coupling by calculating the relative angle $\alpha$ between the orbital direction $\boldsymbol{L}$ of the outgoing state and the $\boldsymbol{g_0}$-vector (Fig. \ref{fig:2}b). The results are reported in Fig. \ref{fig:2}c,d for an input state with $p_x$ and $p_z$ type ($p_y$ yields behavior analogous to that for $p_x$)
and for a given position $z^*$ in 
$\text{III}$. 
The orbital moment filtering ensures that the $\boldsymbol{g_0}$-vector and 
$\boldsymbol{L}$ are never collinear except at specific points, e.g., when a single component of the $\boldsymbol{g}$-vector coupling is considered (Figs. \ref{fig:2}c,d). Indeed, if all mirror and twofold rotational symmetries are broken in centrosymmetric region $\text{II}$, the orbital moment of the transmitted electron can change from a collinear to an anticollinear configuration by adjusting either the out-of-plane ($g_{0z}$) or in-plane orbital couplings ($g_{0x}$, $g_{0y}$) in the parameters space. This leads to scenarios where the orbital moment is perpendicular to the $\boldsymbol{g_0}$-vector. Some of these physical configurations have a distinct profile in the $\boldsymbol{g_0}$-parameters space highlighted by solid lines in Figs. \ref{fig:2}c,d. The corresponding lines are robust to a change in the amplitude of the model parameters far as the mirror symmetries in the orbital space $\hat{M}_{yz}$ and $\hat{M}_{\overline{yz}}$ hold (see Appendix \ref{app:D} for details). 

\begin{figure}[!t]
\centering
\includegraphics[width=0.9\linewidth]{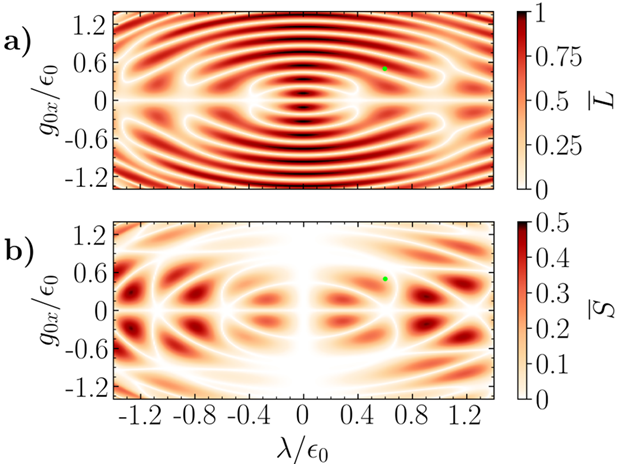}
\caption{
Average orbital (a) and spin (b) moment across the two spin channels at a specific point in the region $\text{III}$ (i.e. $z^*$ in Fig. \ref{fig:3}) within
\textcolor{black}{($g_{0x}$, $\lambda$)} phase space.
Electrons are prepared in a $p_z$ orbital state as an equal mixture of spin up and down configurations aligned along the $z$-axis. Here, $g_{0y}=g_{0z}=0$, $d=20$ nm, $\epsilon_0=38$ meV, and the green point corresponds to the values reported in Fig. \ref{fig:3}e,f. The profiles of the filtered orbital and spin moments within the ($g_{0x}$, $\lambda$) phase space indicate that the peak amplitudes for the orbital and spin channels arise from distinct configurations.
}
\label{fig:4}
\end{figure}

Let us then consider the influence of atomic SOC in the region {II} expressed through the term $ \lambda\hat{\boldsymbol{L}}\cdot\hat{\boldsymbol{S}}$. 
The introduction of the SOC
does not alter the symmetry characteristics of the transmission medium. 
We analyze the electron transmission problem by considering input states that are spin polarized either up or down with respect to a specified direction.
In a manner similar to orbital momentum filtering, we then evaluate the outgoing states and the probabilities for each spin channel to be transferred through the region $\text{II}$. 
The findings are illustrated in Fig. \ref{fig:3} for a specific value of $\lambda$, under the assumption that only one component of the \textcolor{black}{$\boldsymbol{g_0}$}-coupling is active. In this configuration, the presence of SOC results in non-zero values for both spin and orbital projected polarization, regardless of the orientation (Fig. \ref{fig:3}a,b,c,d). By averaging the amplitude of the orbital and spin polarization, respectively, in each direction and taking into account an unpolarized mixture of spin channels for the injected electrons, we observe that the resulting filtering is effective solely for $L_x$ (Fig. \ref{fig:3}e) and $S_x$ (Fig. \ref{fig:3}f). This is expected since the orbital moment filtering is able to polarize only one component 
and, due to the structure of the SOC, the corresponding spin polarization gets activated too. 

We then examine the quantitative behavior related to spin and orbital moment filtering. To achieve this, we investigate the parameter space of the \textcolor{black}{$\boldsymbol{g_0}$}-orbital coupling and the spin-orbit interaction $\lambda$ (see Fig. \ref{fig:4}). For simplicity, we focus solely on the single-component scenario for the $\boldsymbol{g_0}$-coupling. The breaking of all mirror and rotational symmetries in the region $\text{II}$ primarily affects the orientation rather than the amplitude of the filtering effect. 
Furthermore, we examine a particular parameter range for $\lambda$ and $\boldsymbol{g_0}$, with their amplitudes reaching approximately 50 meV for a fixed transmission region size. Changes in the transmission region size or the Fermi energy does not alter the qualitative outcome as it effectively leads to a rescaling of these parameters.
Moreover, we consider a specific parameters range of $\lambda$ and $\boldsymbol{g_0}$ that vary up to an amplitude of about 50 meV for a given size of the transmission region. A variation of the size or of the Fermi energy results into an effective rescaling of the parameters.
\textcolor{black}{We observe that the orbital moment of the outgoing states can be fully polarized even for small values of the symmetry-breaking coupling, $\boldsymbol{g_0}$, provided the atomic spin-orbit coupling is negligible (Fig. \ref{fig:4}a). Typically, spin-orbit coupling influences the amplitude of the orbital moment and can induce a non-zero spin polarization filtering (Fig. \ref{fig:4}b). 
Our analysis shows that by combining the symmetry-breaking coupling with spin-orbit interaction one can achieve maximum spin filtering efficiency even when the spin-orbit coupling $\lambda$ is relatively weak (i.e. around 8 meV). The overall trend of the orbital and spin filtering indicates that, due to interference effects between transferred and reflected electrons at the interface between regions $\text{II}$ and $\text{III}$, there can be outgoing states in $\text{III}$ with minimal spin and orbital moment filtering. We note that these configurations in the parameters space arise only for Fermi wavelengths that are smaller or comparable than the size of the region $\text{II}$.}

\textcolor{black}{Finally, we have analyzed the cases with modified boundary
conditions at the interface between the various regions forming the transmission line. In particular, we consider the case with a delta function potential at the interface so that
there are no continuity conditions for the derivative of the wave functions. The outcome of this analysis indicates that the interface potential can modulate the amplitude of the filtering while it
does not alter the qualitative results regarding the orientation of the spin and orbital filtering (see Appendix \ref{app:C}).}


\textcolor{black}{We would also like to provide a connection with a physical setup that aligns with a specific geometry or crystalline arrangement of the areas constituting the transmission line. To achieve this, one can start considering a basic configuration that can be connected with the examined one dimensional transmission line.
Specifically, we can introduce three regions sharing the same lattice crystalline structure, e.g. with a square geometry, though featuring a misalignment in the axis orientation between regions ($\text{I}$, $\text{III}$) and region $\text{I}$ (see Appendix \ref{app:A}). By utilizing a multi-orbital tight-binding model and distinguishing between the degrees of freedom associated with crystal wave vectors that are parallel and perpendicular to the transmission direction, we can formulate an effective one-dimensional model akin to the one used in the analysis. This can be accomplished by projecting out the transverse degrees of freedom, which applies, for instance, to the case of a thin geometry where the length of the system along the transmission line exceeds the other dimensions.
In particular, the misalignment of the crystalline axes between the region $\text{I}$ and $\text{II}$ is responsible for the lack of common mirrors that, in turn, leads to the orbital couplings employed in the model analysis. This derivation indicates that orbital moment filtering can be achieved with the absence of shared mirror
and rotational symmetry transformations between the injection and transmission regions and, thus, also obtained by employing domains having the same crystalline symmetry. Specifically,
regions $\text{I}$ and $\text{II}$ may exhibit a high-symmetry crystalline configuration, such as possessing multiple vertical mirrors
or twofold rotational symmetries, provided that the associated mirror planes and rotational axes in the
two regions are not equivalent.}

\section{Conclusions}\label{sec:IV}

In conclusion, our results demonstrate that the breaking of mirror and rotational symmetries—rather than inversion symmetry—is essential for effective spin and orbital filtering, as well as for controlling the orientation of the transmitted angular momentum. 
Regarding the phenomenon of chiral-induced spin selectivity, the uncovered mechanisms indicate that these effects can also manifest in achiral molecules that possess inversion symmetry along with at least one \textcolor{black}{common mirror plane and a twofold rotational symmetry axis with respect to the injection region}. 

A direct application of our prediction is provided by the use of vanadocene or similar types of molecules \cite{Pal2019}. 
\textcolor{black}{Indeed, vanadocene possesses $D_{5d}$ point group symmetry, which means that it does not have a horizontal mirror plane and features five vertical mirror planes associated with its planar pentagonal structure. Hence, when it is put in contact with a cubic system, as for the case of elemental Ag, there can be no more than one common vertical mirror depending on the relative orientation of the symmetry axes. Then, the electronic states of the vanadocene will act as a source of orbital hybridization for the Ag states with different mirror parity, thus leading to couplings of the type employed in the examined model.}
Additionally, our investigation of the role of atomic SOC reveals that, when examining electrical transport with metallic leads connected to the transmission medium, it is not essential for the leads to consist of heavy elements with substantial SOC \cite{Liu2021,Adhikari2023,Gersten2013}. 
The capability for spin and orbital filtering does not depend on SOC in the areas where charge current is injected or collected, consistently with recent experimental observations \cite{Hannah2023}.

These findings extend beyond molecules, \textcolor{black}{applying to solid state centrosymmetric materials 
or to quasi one-dimensional semiconducting nanostructures \cite{Streda2003} with electronic states marked by multiple orbital degrees of freedom.
In this context, it can be estimated that with a Fermi wavelength between a few and several tens of nanometers, and an amplitude for the symmetry-breaking term on the order of a few meV, a region II, with length of approximately 50 nanometers, can lead to a maximum orbital polarization filtering of around 70$\%$ and a spin polarization of about 25$\%$ \textcolor{black}{when considering materials with light or heavy elements whose effective atomic spin-orbit splitting amplitudes, close to the center of the Brillouin zone, range from 10 to 100 meV.}
}


\noindent{\bf{Acknowledgements}}\\
L.J.D'O. and M.C. acknowledge support from
PNRR MUR project PE0000023-NQSTI. C.O. and M.T.M. acknowledge partial support by PNRR MUR project PE0000023-NQSTI (TOPQIN) and by the Italian Ministry of Foreign Affairs and International Cooperation, Grants No. PGR12351 (ULTRAQMAT). 
M.C. acknowledges support by Italian Ministry of University and Research (MUR) PRIN 2022 under the Grant No. 2022LP5K7 (BEAT).
F.M. acknowledges the PNRR project NFFA-DI, IR0000015.
W.B. and A.K. acknowledge support by Narodowe Centrum Nauki (NCN, National Science Centre, Poland) Project No. 2019/34/E/ST3/00404 and by the Foundation for Polish Science project “MagTop” no. FENG.02.01IP.05-0028/23 co-financed by the European Union from the funds of Priority 2 of the European Funds for a Smart Economy Program 2021–2027 (FENG).
We acknowledge valuable discussions with Jeroen van den Brink, Panagiotis Kotetes and Stefano Carretta.

\begin{figure*}
\centering
\includegraphics[width=0.8\textwidth]{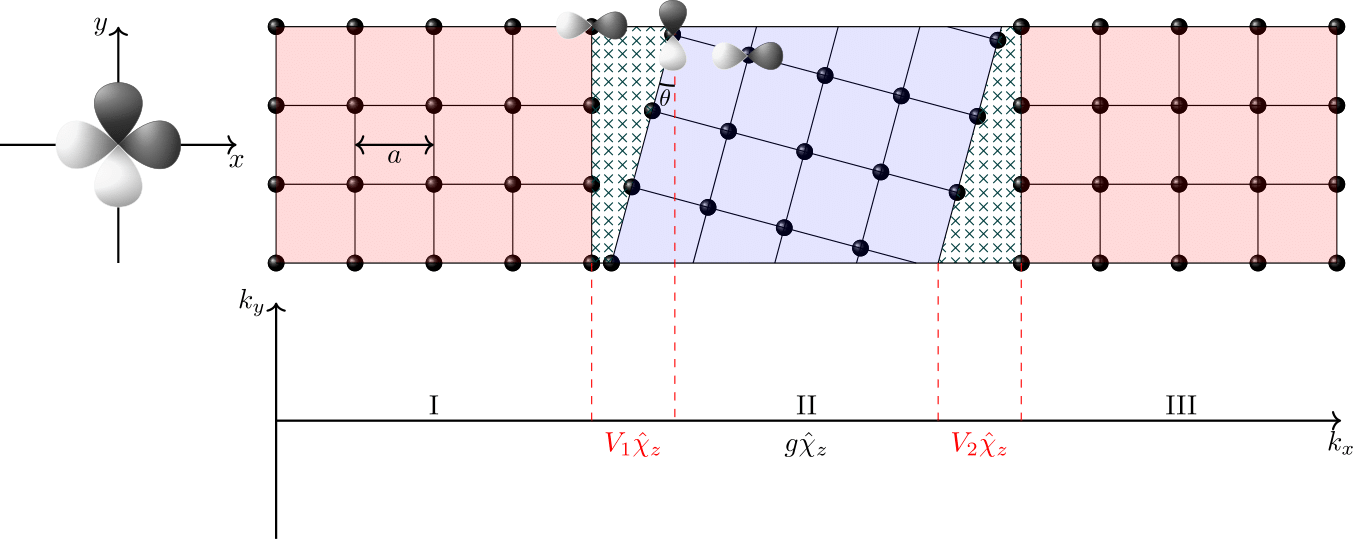}
\textcolor{black}{\caption{This schematic depicts a setup composed of regions having a square lattice geometry, featuring misaligned axes with a given angle $\theta$ among the regions ($\text{I}$, $\text{III}$) with respect to region $\text{II}$. The mismatch in axis orientation allows for the hybridization of mirror-symmetric orbital configurations, specifically $p_x$ and $p_y$, both at the interface, through the $V_1$ and $V_2$ couplings, and within region $\text{II}$ by the term $g$. Consequently, the orbital couplings represented by $\hat{\chi}_z = \hat{L}_x \hat{L}_y+\hat{L}_y\hat{L}_x$ can become non-zero at both interface and in region $\text{II}$.}
\label{suppl_fig:1}}
\end{figure*}

\appendix
\textcolor{black}{
\section{Derivation of the effective model with crystalline arrangement of the transmission line domains}\label{app:A}
In this section, we provide a connection between the examined model in the main text and a physical setup that aligns with a specific geometry or crystalline arrangement of the areas constituting the transmission line.
To this aim, we consider the transmission problem for an electron propagating through three square-lattice regions: $\text{I}$, $\text{II}$, and $\text{III}$, each with the same lattice constant $a$ (see Fig. \ref{suppl_fig:1}).
We assume there is only one atom per unit cell in each region. \textcolor{black}{In addition, within a given reference frame, we consider the system to be strongly confined along the $y$-direction, $d_y \ll d_x$, such that the electronic transmission takes place along the $x$-direction.} 
The crystallographic axes of the outer lattice regions ($\text{I}$ and $\text{III}$) are aligned with the reference frame, while the ones of the intermediate region ($\text{II}$) are rotated by an angle $\theta$, as depicted in Fig. \ref{suppl_fig:1}. By adopting a tight-binding approach, we consider a multi-orbital system whose electronic states at a given energy $\mu$ are
expressed in terms of the atomic orbitals $|p_z\rangle$, $|p_y\rangle$, and $|p_x\rangle$ that span the $L=1$ manifold. All the energies are given in unit of $\epsilon_{0}=\hbar^{2}k_{0}^{2}/2m$, with
$m$ being the mass of the electron. From here on, $\hbar=1$ and $2m=1$ for convenience and clarity.}
\textcolor{black}{By exploiting the hypotheses by Slater and Koster \cite{PhysRev.94.1498} within the first-neighbor approximation, the momentum-space Hamiltonian in region $\text{I}$ and $\text{III}$ is diagonal in the orbital basis. On the other hand, being the crystallographic axes of region $\text{II}$ tilted with respect to the $p$-basis, hybridization effects due to non-vanishing $p_{x}-p_{y}$ overlap emerge both at the interfaces and within the region itself. In particular, it can be shown that
\textcolor{black}{\begin{equation}
\hat{\mathcal{H}}^{\text{II}}\left(\mathbf{k}\right)=\mathcal{\hat{T}}_{0}+2\cos\left(\mathbf{k}\cdot\boldsymbol{a}_{1}\right)\mathcal{\hat{T}}_{1}+2\cos\left(\mathbf{k}\cdot\boldsymbol{a}_{2}\right)\mathcal{\hat{T}}_{2},
\end{equation}
\noindent where
\begin{equation}
\begin{cases}
\mathcal{\hat{T}}_{0}=\xi_{x}^{\text{II}}\hat{L}_{x}^{2}+\xi_{y}^{\text{II}}\hat{L}_{y}^{2}+\xi_{z}^{\text{II}}\hat{L}_{z}^{2}\\
\mathcal{\hat{T}}_{1}=\gamma_{x}^{\text{II}}\hat{L}_{x}^{2}+\gamma_{y}^{\text{II}}\hat{L}_{y}^{2}+\gamma_{z}^{\text{II}}\hat{L}_{z}^{2}+g_{1}\hat{\chi}_{z}\\
\mathcal{\hat{T}}_{2}=\delta_{x}^{\text{II}}\hat{L}_{x}^{2}+\delta_{y}^{\text{II}}\hat{L}_{y}^{2}+\delta_{z}^{\text{II}}\hat{L}_{z}^{2}+g_{2}\hat{\chi}_{z}
\end{cases}
\end{equation}
with $\hat{\chi}_{z}=\hat{L}_{x}\hat{L}_{y}+\hat{L}_{y}\hat{L}_{x}$. Here, the terms proportional to $\hat{L}_{i}^{2}$
reproduce symmetry allowed crystal-field effects; on the contrary, $\hat{\chi}_z$ preserves inversion $\hat{\mathcal{I}}:\left(x,y,z\right)\rightarrow\left(-x,-y,-z\right)$, while breaks
$\hat{\mathcal{M}}_{x,y}$ and $\hat{\mathcal{C}}_{2x,2y}$ by coupling the electronic orbital degrees of freedom. The involved parameters ($\gamma$, $\delta$, $\xi$, and $g$) depend on the angle $\theta$, the
on-site energies, and the two-center bond integrals between two $p$-orbitals, $V_{pp\sigma}$ and $V_{pp\pi}$, where $\sigma$ and $\pi$ denote the type of the chemical bond \cite{PhysRev.94.1498}. Particularly, $g_{1,2}=0$ if $\theta=0$. 
In the long-wavelength approximation, the trigonometric functions can be expanded as $\cos x\sim1-x^{2}/2$ and $\sin x\sim x$. Thus, by retaining only terms quadratic in momentum, the effective Hamiltonian in the continuous limit takes the form 
\begin{equation}
\hat{\mathcal{H}}^{\text{II}}=\hat{A}\hat{p}_{x}^{2}+\hat{B}\hat{p}_{y}^{2}+\hat{C}\hat{p}_{x}\hat{p}_{y}+\hat{D},
\end{equation}
\noindent where $\hat{p}_{i}$ is the electronic linear momentum operator along the $i$-axis, and
\begin{equation}
\begin{cases}
\hat{A}=-a^{2}\cos^{2}\theta\mathcal{\hat{T}}_{1}-a^{2}\sin^{2}\theta\mathcal{\hat{T}}_{2}\\
\hat{B}=-a^{2}\sin^{2}\theta\mathcal{\hat{T}}_{1}-a^{2}\cos^{2}\theta\mathcal{\hat{T}}_{2}\\
\hat{C}=2a^{2}\sin\theta\cos\theta\left(\mathcal{\hat{T}}_{1}-\mathcal{\hat{T}}_{2}\right)\\
\hat{D}=\mathcal{\hat{T}}_{0}+2\mathcal{\hat{T}}_{1}+2\mathcal{\hat{T}}_{2}
\end{cases}.
\end{equation}
We choose a complete basis composed of factorized states, $\left\{ \left|\Psi_{n}\left(x,y\right)\right\rangle =\psi_{n_{1}}\left(x\right)\phi_{n_{2}}\left(y\right)\left|\eta_{n_{3}}\right\rangle \right\}$,  such that $\left\langle \psi_{n_{1}}|\psi_{m_{1}}\right\rangle =\delta_{n_{1},m_{1}}$, $\left\langle \phi_{n_{2}}|\phi_{m_{2}}\right\rangle =\delta_{n_{2},m_{2}}$, and $\left\langle \eta_{n_{3}}|\eta_{m_{3}}\right\rangle =\delta_{n_{3},m_{3}}$, with  $|\eta_{n_{3}}\rangle$ being the basis associated with the orbital degrees of freedom. Given the confinement of the system along the $y$-direction, we impose hard-wall boundary conditions along this axis: $\left|\Psi\left(x,y=\pm d_{y}/2\right)\right\rangle =0$. This leads to considering 
\begin{equation}
\phi_{n}^{+}\left(y\right)=d_{y}^{-1/2}\cos\left(k_{y,n}^{+}y\right),\,\,\,\phi_{n}^{-}\left(y\right)=d_{y}^{-1/2}\sin\left(k_{y,n}^{-}y\right)
\end{equation}
as electronic transverse modes, where $+$ ($-$) denotes even (odd) parity and
\begin{equation}
k_{y,n}^{+}=\frac{\left(2n-1\right)\pi}{d_{y}},\,\,\,k_{y,n}^{-}=\frac{2n\pi}{d_{y}}
\end{equation}
with $n=1,2,\ldots$. Applying the operator $\hat{p}_{y}^{2}$ on the transverse modes yields the corresponding confinement energies
\begin{equation}
E_{y,n}^{+}=\frac{\left(2n-1\right)^{2}\pi^{2}}{d_{y}^{2}},\,\,\,
E_{y,n}^{-}=\frac{4n^{2}\pi^{2}}{d_{y}^{2}};
\end{equation}
here, the lowest-energy sub-band due to confinement along $y$ is $\pi^{2}/d_{y}^{2}$, with an energy gap of $3\pi^{2}/d_{y}^{2}$ to the next sub-band.
Then, the matrix representation of the Hamiltonian in the chosen basis is given by $\hat{h}_{n,m}=\left\langle \Psi_{n}\left|\hat{\mathcal{H}}\right|\Psi_{m}\right\rangle$. It follows that
\begin{align}
\hat{h}_{n,m} &= \left\langle \psi_{n_{1}} \left| \hat{p}_{x}^{2} \right| \psi_{m_{1}} \right\rangle \delta_{n_{2},m_{2}} \hat{A}_{n_{3},m_{3}} + \nonumber \\ &\quad \delta_{n_{1},m_{1}} \delta_{n_{2},m_{2}} E_{y,m_{2}}\hat{B}_{n_{3},m_{3}} + \nonumber \\
&\quad \left\langle \psi_{n_{1}} \left| \hat{p}_{x} \right| \psi_{m_{1}} \right\rangle \left\langle \phi_{n_{2}} \left| \hat{p}_{y} \right| \phi_{m_{2}} \right\rangle \hat{C}_{n_{3},m_{3}} + \nonumber \\ 
&\quad \delta_{n_{1},m_{1}} \delta_{n_{2},m_{2}} \hat{D}_{n_{3},m_{3}}.
\end{align}
It can be shown that the matrix element $\left\langle \phi_{n_{2}} \left| \hat{p}_{y} \right| \phi_{m_{2}} \right\rangle$ vanishes when $\phi_{n_{2}}$ and $\phi_{m_{2}}$ have the same parity (i.e., both even or both odd), and can be nonzero only when their parities differ. Therefore, $\hat{h}$ assumes the following block structure:
\begin{equation}
\hat{h}=\left(\begin{array}{ccc}
\hat{h}_{1D}^{\left(1\right)} & \hat{h}_{c}^{\left(1,2\right)}\\
\hat{h}_{c}^{\left(2,1\right)} & \hat{h}_{1D}^{\left(2\right)}\\
 &  & \ddots
\end{array}\right).
\end{equation}
The $i$-th diagonal block corresponds to the one-dimensional effective Hamiltonian for the $i$-th sub-band, namely
\begin{equation}
\hat{h}_{1D}^{\left(i\right)}=\hat{A}\hat{p}_{x}^{2}+\hat{B}E_{y,i}+\hat{D}.
\end{equation}
On the other hand, the off-diagonal block
\begin{equation}\hat{h}_{c}^{\left(i,j\right)}=\left\langle \phi_{i}\left|\hat{p}_{y}\right|\phi_{j}\right\rangle \hat{C}\hat{p}_{x}
\end{equation}
accounts for the coupling (in case of different parities) between the $i$-th and $j$-th sub-bands mediated by transverse mode mixing.}}

\textcolor{black}{Crystal axis misalignment can result also in a $y$-dependent variation of the hopping amplitude at the interfaces. This physical scenario can be modeled by introducing a barrier potential acting at the left (L) and right (R) interfaces, expressed as 
\begin{equation}
f\left(x,y\right)=\hat{E}V_{L}\left(y\right)\delta\left(x+d_{x}/2\right)+\hat{F}V_{R}\left(y\right)\delta\left(x-d_{x}/2\right)
\end{equation}
with $\hat{E}=\left[\sum_{i}\nu_{L,i}\hat{L}_{i}^{2}+\tilde{g}_{L}\hat{\chi}_{z}\right]$,
$\hat{F}=\left[\sum_{i}\nu_{R,i}\hat{L}_{i}^{2}+\tilde{g}_{R}\hat{\chi}_{z}\right]$, and the parameters $\nu$ and $\tilde{g}$ depending on the electronic and geometrical properties of the interfaces. Here, $V_{L}$ and $V_{R}$ are $y$-dependent barrier profiles that attain large amplitudes where the atomic separation suppresses the transmission probability. In this
framework, using the $\phi_{n}$ representation for the transverse
modes, we can rewrite the Hamiltonian in a way that each block (for
the intra- and inter-band contributions) will include delta-function potentials along the $x$-direction depending on the character of the involved transverse modes. In particular, we have that the $i$-th diagonal block becomes 
\begin{align}\hat{h}_{1D}^{\left(i\right)}= & \hat{A}\hat{p}_{x}^{2}+\hat{B}E_{y,i}+\hat{D}+\hat{E}\tilde{V}_{L,i}\delta\left(x+d_{x}/2\right)+ \nonumber \\
 & \hat{F}\tilde{V}_{R,i}\delta\left(x-d_{x}/2\right),
\end{align}
where $\tilde{V}_{L,i}=\left\langle \phi_{i}\left|V_{L}\left(y\right)\right|\phi_{i}\right\rangle $
($\tilde{V}_{R,i}=\left\langle \phi_{i}\left|V_{R}\left(y\right)\right|\phi_{i}\right\rangle $) corresponds to the effective amplitude of a scattering delta potential present
at the left (right) interface. At the same time, the off-diagonal
block coupling the $i$-th and $j$-th mode takes the form
\begin{align}
\hat{h}_{c}^{\left(i,j\right)}=&\left\langle \phi_{i}\left|\hat{p}_{y}\right|\phi_{j}\right\rangle \hat{C}\hat{p}_{x}+\hat{E}\tilde{V}_{L,ij}\delta\left(x+d_{x}/2\right)+\nonumber\\&\hat{F}\tilde{V}_{R,ij}\delta\left(x-d_{x}/2\right),
\end{align}
where $\tilde{V}_{L,ij}=\left\langle \phi_{i}\left|V_{L}\left(y\right)\right|\phi_{j}\right\rangle $
and $\tilde{V}_{R,ij}=\left\langle \phi_{i}\left|V_{R}\left(y\right)\right|\phi_{j}\right\rangle $.
At this stage, we exploit the hypothesis of ultra-thin limit regime by considering only the lowest-energy sub-band associated with confinement along $y$. This approximation is well-suited, as the energy separation between the ground and first excited sub-band scales as $d_{y}^{-2}$ and, therefore, is significantly large. It corresponds to set $n_{2}=m_{2}=1$, leading to an effective Hamiltonian for the system of the form
\begin{align}
\hat{h}_{eff}=&\hat{A}\hat{p}_{x}^{2}+\hat{B}E_{y,1}+\hat{D}+\hat{E}\tilde{V}_{L,1}\delta\left(x+d_{x}/2\right)+\nonumber\\&\hat{F}\tilde{V}_{R,1}\delta\left(x-d_{x}/2\right).
\end{align}
It is worth noting that, even in the limit of having only the lowest quantum well state occupied at the Fermi level with ideal interfaces ($\tilde{V}_{L,1}$=$\tilde{V}_{R,1}=0$), the effects of coupling between the confined states and the channel emerge through the term $\hat{B}E_{y,1}$.
We also observe that each block-diagonal Hamiltonian shares identical symmetry-breaking terms linked to the $\hat{\chi}_z$ component. Then, the study of the effective 1D model Hamiltonian can capture the essential physics of the spin and orbital filtering effects in the transport of thin nanostructures. Consequently, the resulting effective Hamiltonian for the system is
\begin{equation}
\begin{cases}
\hat{\mathcal{H}}^{\text{I}}=\hat{p}_{x}^{2}\left(\sum_{i}a_{i}^{\text{I}}\hat{L}_{i}^{2}\right)+\sum_{i}\Delta_{i}^{\text{I}}\hat{L}_{i}^{2}\\\hat{\mathcal{H}}^{\text{II}}=\hat{p}_{x}^{2}\left(\sum_{i}a_{i}^{\text{II}}\hat{L}_{i}^{2}+g_{z}\hat{\chi}_{z}\right)+\sum_{i}\Delta_{i}^{\text{II}}\hat{L}_{i}^{2}+\\\,\,\,\,\,\,\,\,\,\,\,\,\,\,\,\,g_{0z}\hat{\chi}_{z}+\hat{V}\left(x\right)\\\hat{\mathcal{H}}^{\text{III}}=\hat{p}_{x}^{2}\left(\sum_{i}a_{i}^{\text{III}}\hat{L}_{i}^{2}\right)+\sum_{i}\Delta_{i}^{\text{III}}\hat{L}_{i}^{2}
\end{cases},
\end{equation}
\noindent where
\begin{equation}
\begin{cases}
\hat{V}\left(x\right)=V_{1}\delta\left(x+d_{x}/2\right)\hat{E}+V_{2}\delta\left(x-d_{x}/2\right)\hat{F}\\
a_{i}^{\text{II}}=-a^{2}\left(\gamma_{i}^{\text{II}}\cos^{2}\theta+\delta_{i}^{\text{II}}\sin^{2}\theta\right)\\
g_{z}=-a^{2}\left(g_{1}\cos^{2}\theta+g_{2}\sin^{2}\theta\right)\\
\Delta_{i}^{\text{II}}=\alpha_{y}\gamma_{i}^{\text{II}}+\beta_{y}\delta_{i}^{\text{II}}+\xi_{i}^{\text{II}}\\
g_{0z}=\alpha_{y}g_{1}+\beta_{y}g_{2}
\end{cases}
\end{equation}
with $\alpha_{y}=2-a^{2}E_{y,1}\sin^{2}\theta$ and $\beta_{y}=2-a^{2}E_{y,1}\cos^{2}\theta$. 
Further aspects related to this approach for the study of heterojunctions can be found in Refs. \cite{Ivchenko1997,vasko1998electronic}.}

\textcolor{black}{
Hereafter, for clarity with respect to the notation in the main text, we assume that the electron propagates along the $z$-axis of a coordinate system $Oxyz$. 
To include the electronic spin degree, we employ the Pauli matrices: $\hat{S}_{i=x,y,z}=\frac{1}{2}\hat{\sigma}_{i=x,y,z}$. For the analysis of the transmission problem, the electron is prepared in a given state $\left|\Psi_{in}^{\text{I}}\right\rangle$ in region $\text{I}$. There, the Hamiltonian is expressed as $\hat{\mathcal{H}}^{\text{I}}\left(z\right)=\left[\hat{p}_{z}^{2}\left(\sum_{i}a_{i}^{\text{I}}\hat{L}_{i}^{2}\right)+\sum_{i}\Delta_{i}^{\text{I}}\hat{L}_{i}^{2}\right]\hat{\sigma}_{0}$, with $\hat{\sigma}_{0}=\hat{\sigma}_{i}^{2}$. The electron is thus injected
from region $\text{I}$ into region $\text{II}$, which spans from
$z=-d/2$ to $z=d/2$. \textcolor{black}{In this region, considering the factors involved in deriving the effective model for the mismatched heterostructure—taking into account various geometries or crystalline symmetries—the single-electron Hamiltonian will include all the components of the $\chi$-coupling, thus having the following expression}:
\textcolor{black}{
\begin{align}
\hat{\mathcal{H}}^{\text{II}}(z)&=\left[\hat{p}_{z}^{2}\left(\sum_{i}a_{i}^{\text{II}}\hat{L}_{i}^{2}+\boldsymbol{g}\cdot\hat{\boldsymbol{\chi}}\right)\right]\hat{\sigma}_{0}+\nonumber\\&\quad\left(\sum_{i}\Delta_{i}^{\text{II}}\hat{L}_{i}^{2}+\boldsymbol{g_{0}}\cdot\hat{\boldsymbol{\chi}}\right)\hat{\sigma}_{0}+\lambda\hat{\boldsymbol{L}}\cdot\hat{\boldsymbol{S}}+\nonumber\\&\quad\left[V_{1}\delta\left(z+d/2\right)\hat{\mathcal{V}}_{1}+V_{2}\delta\left(z-d/2\right)\hat{\mathcal{V}}_{2}\right]\hat{\sigma}_{0}
\end{align}
with $\boldsymbol{g}=\left(g_{x},g_{y},g_{z}\right)$, $\boldsymbol{g_{0}}=\left(g_{0x},g_{0y},g_{0z}\right)$, $\hat{\boldsymbol{\chi}}=\left(\left\{ \hat{L}_{y},\hat{L}_{z}\right\} ,\left\{ \hat{L}_{x},\hat{L}_{z}\right\} ,\left\{ \hat{L}_{x},\hat{L}_{y}\right\} \right)$, $\lambda$ as the atomic spin-orbit coupling magnitude (in units of $\epsilon_{0}$), and $\hat{\mathcal{V}}_{i}=\sum_{j}\left(\nu_{i,j}\hat{L}_{j}^{2}+\tilde{g}_{i,j}\hat{\chi}_{j}\right)$.} 
Finally, the electron can reach the last region of space, region $\text{III}$,
where its orbital and spin angular momenta are measured,
and the Hamiltonian can be expressed as $\hat{\mathcal{H}}^{\text{III}}\left(z\right)=\left[\hat{p}_{z}^{2}\left(\sum_{i}a_{i}^{\text{III}}\hat{L}_{i}^{2}\right)+\sum_{i}\Delta_{i}^{\text{III}}\hat{L}_{i}^{2}\right]\hat{\sigma}_{0}$.
Thus, we can define the following overall Hamiltonian $\hat{\mathcal{H}}=\hat{\mathcal{H}}\left(z\right)$:
\begin{equation}
\hat{\mathcal{H}}\left(z\right)=\begin{cases}
\hat{\mathcal{H}}^{\text{I}}\left(z\right) & z<-d/2\\
\hat{\mathcal{H}}^{\text{II}}\left(z\right) & -d/2\leq z\leq d/2\\
\hat{\mathcal{H}}^{\text{III}}\left(z\right) & z>d/2
\end{cases},
\end{equation}
where $\mathcal{\hat{H}}^{A}\left(z\right)=\hat{\mathcal{A}}^{A}\hat{p}_{z}^{2}+\hat{\mathcal{B}}^{A}$ for $A=\text{I},\text{II},\text{III}$.}

\section{Solving the transmission problem}\label{app:B}
To access the orbital and spin angular moment components of the
transmitted electron in region $\text{III}$, we need to know the electronic
state there. The latter can be determined by solving the following
quantum mechanics problem. Let indicate with $\left|\Psi^{\text{I}}\right\rangle $, $\left|\Psi^{\text{II}}\right\rangle $,
and $\left|\Psi^{\text{III}}\right\rangle $ the electronic state in region
$\text{I}$, $\text{II}$, and $\text{III}$, respectively. Each of them has to satisfy the corresponding
(stationary) Schrödinger equation, i.e., $\hat{\mathcal{H}}^{A}\left|\Psi^{A}\right\rangle =\mu\left|\Psi^{A}\right\rangle $
with $A=\text{I},\text{II},\text{III}$. Thus, in light of the Hamiltonian of the system,
it can be proved that the most generic way of writing the electronic
state in any region is
\begin{equation}
\left|\Psi^{A}\left(z\right)\right\rangle =\sum_{j=1}^{6}\left(\alpha_{j}^{A}e^{ik_{j}^{A}z}+\beta_{j}^{A}e^{-ik_{j}^{A}z}\right)\left|\phi_{j}^{A}\right\rangle ,
\end{equation}

i.e., the linear combination of transmitted and reflected plane waves
with the electronic orbital and spin character contained in the vector
$\left|\phi_{j}^{A}\right\rangle $. Here, the generic electron wavenumber $k_{j}^{A}$ is expressed in units of $k_{0}$. In our case, we put $\left|\Psi_{in}^{\text{I}}\left(z\right)\right\rangle =\sum_{j=1}^{6}\alpha_{j}^{\text{I}}e^{ik_{j}^{\text{I}}z}\left|\phi_{j}^{\text{I}}\right\rangle $;
in addition, assuming there is only a transmitted wavefunction in
region $\text{III}$, we pose $\beta_{j=1,\ldots,6}^{\text{III}}=0$. The state has
to be continuous in space, as well as its first derivative \textcolor{black}{when $\hat{\mathcal{A}}^{\text{I}}=\hat{\mathcal{A}}^{\text{II}}=\hat{\mathcal{A}}^{\text{III}}$ and there are no singularities in space ($V_1=V_2=0$)}. Therefore,
the following conditions have to be fulfilled simultaneously:
\begin{equation}
\begin{cases}
\left|\Psi^{\text{I}}\left(-d/2\right)\right\rangle =\left|\Psi^{\text{II}}\left(-d/2\right)\right\rangle \\
\left|\Psi^{	\text{II}}\left(d/2\right)\right\rangle =\left|\Psi^{\text{III}}\left(d/2\right)\right\rangle \\
{D}\left|\Psi^{\text{I}}\left(-d/2\right)\right\rangle ={D}\left|\Psi^{\text{II}}\left(-d/2\right)\right\rangle \\
{D}\left|\Psi^{	\text{II}}\left(d/2\right)\right\rangle ={D}\left|\Psi^{\text{III}}\left(d/2\right)\right\rangle \\
\end{cases},
\end{equation}
where $D\left|\Psi^{A}\left(\bar{z}\right)\right\rangle =\left.\left(d\left|\Psi^{A}\left(z\right)\right\rangle /dz\right)\right|_{z=\bar{z}}$.
In doing so, we obtain a system of 24 linear equations with 24 variables
($\beta_{1,\ldots,6}^{\text{I}},\alpha_{1,\ldots,6}^{	\text{II}},\beta_{1,\ldots,6}^{	\text{II}},\alpha_{1,\ldots,6}^{\text{III}}$). Thus, by solving it, it is possible to construct the electronic state
in each region. In particular, the outcoming electronic state, i.e., the one in region $\text{III}$, can be written in terms of $\left|\Psi^{\text{I}}_{in}\right\rangle$ through a `transmission' matrix $\hat{\mathcal{T}}$ such that
\begin{equation}
	\left|\Psi^{\text{III}}\left(z\right)\right\rangle =\hat{\mathcal{T}}\left|\Psi^{\text{I}}_{in}\left(z\right)\right\rangle.
\end{equation} 
Furthermore, we can compute the space-dependent single-electron mean
orbital and spin angular moment, defined as follows: 
\begin{equation}
\boldsymbol{\mathcal{O}}^{A}\left(z\right)=\left(\mathcal{O}_{x}^{A}\left(z\right),\mathcal{O}_{y}^{A}\left(z\right),\mathcal{O}_{z}^{A}\left(z\right)\right),
\end{equation}
 with $\mathcal{O}=L$, $S$, respectively, $A=\text{I}$ ($z<-d/2$), $\text{II}$ ($-d/2\leq z\leq d/2$),
$\text{III}$ ($z>d/2$), and
\begin{equation}
\mathcal{O}_{j}^{A}\left(z\right)=\frac{\left\langle \Psi^{A}\left(z\right)\right|\hat{\mathcal{O}_{j}}\left|\Psi^{A}\left(z\right)\right\rangle }{\left\langle \Psi^{A}\left(z\right)\right|\left.\Psi^{A}\left(z\right)\right\rangle }.
\end{equation}

The mean orbital and spin angular moment components in region $\text{III}$
($z>d/2$) cannot generally be determined a priori, as they depend
on the initial configuration of the system and on the parameters defining
its Hamiltonian. However, it is worth noticing that, if the transmission matrix $\hat{\mathcal{T}}$ preserves certain symmetries, it may be possible to extract meaningful
information about those components. Let suppose that $\left[\hat{\mathcal{T}},\hat{\mathcal{U}}\right]=0$,
where $\hat{\mathcal{U}}$ is a generic unitary operator. Being $\hat{\mathcal{T}}=\hat{\mathcal{U}}^{-1}\hat{\mathcal{T}}\hat{\mathcal{U}}$, we have that
\begin{equation}
\hat{\mathcal{U}}\left|\Psi^{\text{III}}\left(z\right)\right\rangle =\hat{\mathcal{T}}\left(\hat{\mathcal{U}}\left|\Psi^{\text{I}}_{in}\left(z\right)\right\rangle \right)
\end{equation}
It means that, if the system is initially prepared in ${\left|\tilde{\Psi}_{in}^{\text{I}}\right\rangle }=\hat{\mathcal{U}}\left|\Psi_{in}^{\text{I}}\right\rangle $, where $\left[\hat{\mathcal{T}},\hat{\mathcal{U}}\right]=0$,
the corresponding electronic state in region $\text{III}$ is simply given by ${\left|\tilde{\Psi}^{\text{III}}\right\rangle }=\hat{\mathcal{U}}\left|\Psi^{\text{III}}\right\rangle $.
As we will see, owing to this, we can gain insight about the properties
of the electronic angular moment. In the following, we denote the electronic state and the mean orbital
moment components in region $\text{III}$ by $\left|\Psi\left(z,\phi_{in}\right)\right\rangle $
and $L_{i}\left(z,\phi_{in}\right)$, respectively, where $i=x,y,z$
and $\left|\phi_{in}\right\rangle $ is the corresponding initial
orbital configuration.
 
\section{Transmission Matrix and representative physical cases}\label{app:C}
In this section we show the solution of the transmission problem by evaluating the transmission matrix that allows to link the initial state with the outgoing state. Additionally, we analyze different configurations assuming nonvanishing crystal field potentials.
Let us focus on few illustrative cases discussed in the main text. \textcolor{black}{We start by considering the following parameters: $a_{i=x,y,z}^{A=\text{I},\text{II},\text{III}}=0.5$, $\boldsymbol{g_0}=\left(g_{0x}\neq0,0,0\right)$,
$\Delta_{i=x,y,z}^{A=\text{I},\text{II},\text{III}}=0$, $\boldsymbol{g}=\boldsymbol{0}$, and $\lambda=0$ (we can, thus, discard the spin degree of freedom).} So, the Hamiltonian reduces to

\begin{equation}
\hat{\mathcal{H}}\left(z\right)=\begin{cases}
\hat{p}_{z}^{2}\text{\ensuremath{\mathcal{I}}} & z<-d/2\\
\hat{p}_{z}^{2}\text{\ensuremath{\mathcal{I}}}+g_{0x}\hat{\chi}_{x} & -d/2\leq z\leq d/2\\
\hat{p}_{z}^{2}\text{\ensuremath{\mathcal{I}}} & z>d/2
\end{cases}, 
\label{eq:ham_gx_0}
\end{equation}

Here, $\mathcal{I}=\left(1/2\right)\left[\hat{L}_{x}^{2}+\hat{L}_{y}^{2}+\hat{L}_{z}^{2}\right]$
is the three-dimensional identity matrix. It follows that $k_{j}^{\text{I}}=k_{j}^{\text{III}}=k$
for any $j=1,2,3$, with $k=\sqrt{\mu}$. It can be proved that $\hat{\mathcal{H}}$ is invariant
under mirroring with respect to the plane orthogonal
to the $x$-axis and
passing through the origin $O$, i.e.,  $\left(x,y,z\right)\rightarrow\left(-x,y,z\right)$: $\left[\hat{\mathcal{H}},\hat{\mathcal{M}}_{x}\right]=0$. $\hat{\mathcal{M}}_{x}$ can be written as the product of a continuous operator $\hat{m}_{x}$, acting on the real-space part of the electronic state, and of the matrix 
\begin{align}
	\hat{M}_{x} & =\left(\begin{array}{ccc}
		1 & 0 & 0\\
		0 & 1 & 0\\
		0 & 0 & -1
	\end{array}\right),
\end{align}
defined in the orbital space spanned by $\left\{ \left|p_{z}\right\rangle ,\left|p_{y}\right\rangle ,\left|p_{x}\right\rangle \right\}$. In addition, $\hat{\mathcal{H}}$ commutes with the symmetry operators $\hat{M}_{yz}$ and $\hat{M}_{\overline{yz}}$ that exchange $\hat{L}_{y}$ with $\hat{L}_{z}$ and $-\hat{L}_{z}$, respectively:
\begin{align}
	\hat{M}_{yz}=\left(\begin{array}{ccc}
		0 & 1 & 0\\
		1 & 0 & 0\\
		0 & 0 & 1
	\end{array}\right),\,\,\,\hat{M}_{\overline{yz}}=\left(\begin{array}{ccc}
		0 & -1 & 0\\
		-1 & 0 & 0\\
		0 & 0 & 1
	\end{array}\right).
\end{align}

In this case, it can be proved that, for any initial electronic configuration, the transmission matrix $\hat{\mathcal{T}}$ assumes the following form:
\begin{equation}
	\hat{\mathcal{T}}=\left(\begin{array}{ccc}
		s_{1} & s_{2} & 0\\
		s_{2} & s_{1} & 0\\
		0 & 0 & 1
	\end{array}\right),
\end{equation}

where $s_1$ and $s_2$ are complex numbers. Thus, it is evident that $\hat{\mathcal{T}}$ commutes with $\hat{M}_{x}$, $\hat{M}_{yz}$, and $\hat{M}_{\overline{yz}}$. We now demonstrate how these symmetry properties can be exploited in the evaluation of the mean angular moment, considering distinct initial orbital configurations.

Firstly, let imagine the electron initially prepared in $\left|\Psi_{in}^{\text{I}}\left(z\right)\right\rangle =e^{ikz}\left|p_{x}\right\rangle$,
eigenstate of $\hat{\mathcal{H}}^{\text{I}}$. Given the form of $\hat{\mathcal{T}}$, we have that $\left|\Psi\left(z,p_{x}\right)\right\rangle =e^{ikz}\left|p_{x}\right\rangle$. It
follows that $L_{i}\left(z,p_{x}\right)=0$ for $i=x$, $y$, $z$.

Then, let consider $\left|\Psi_{in}^{\text{I}}\left(z\right)\right\rangle =e^{ikz}\left|p_{y}\right\rangle$.
In this case, it is easy to show that $L_{i}\left(z,p_{y}\right)=0$
for $i=y$, $z$. Indeed, we have that
\begin{equation} \hat{\mathcal{M}}_{x}e^{ikz}\left|p_{y}\right\rangle=\hat{m}_{x}e^{ikz}\hat{M}_{x}\left|p_{y}\right\rangle=e^{ikz}\left|p_{y}\right\rangle
\end{equation}
and, because $\left[\hat{\mathcal{T}},\hat{M}_{x}\right]=0$, we can surely write
\begin{equation}
\left|\Psi\left(z,p_{y}\right)\right\rangle =\hat{M}_{x}\left|\Psi\left(z,p_{y}\right)\right\rangle.
\end{equation}
Therefore, being $\hat{M}_{x}^{-1}\hat{L}_{i=y,z}\hat{M}_{x}=-\hat{L}_{i=y,z}$,
it emerges that 
\begin{equation}
L_{i=y,z}\left(z,p_{y}\right)=-L_{i=y,z}\left(z,p_{y}\right)
\end{equation}
 and, thus, $L_{i=y,z}\left(z,p_{y}\right)=0$. 
 
Similarly, because $\hat{\mathcal{M}}_{x}e^{ikz}\left|p_{z}\right\rangle=e^{ikz}\left|p_{z}\right\rangle$,
we have that $L_{i=y,z}\left(z,p_{z}\right)=0$. On the other
hand, being $\hat{M}_{x}^{-1}\hat{L}_{x}\hat{M}_{x}=\hat{L}_{x}$,
we cannot reach the same conclusion for the corresponding components
along the $x$-axis, which can be determined by solving the above-mentioned
transmission problem. Nonetheless, by exploiting the fact that $\left[\hat{\mathcal{T}},\hat{M}_{yz}\right]=0$
(or $\left[\hat{\mathcal{T}},\hat{M}_{\overline{yz}}\right]=0$),
we can prove that $L_{x}\left(z,p_{y}\right)=-L_{x}\left(z,p_{z}\right)$.
Indeed, by assuming $\left|\Psi_{in}^{\text{I}}\left(z\right)\right\rangle =e^{ikz}\left|p_{y}\right\rangle$
as the initial configuration, we have that $e^{ikz}\left|p_{y}\right\rangle=\hat{M}_{yz}e^{ikz}\left|p_{z}\right\rangle$
and, therefore, $\left|\Psi\left(z,p_{y}\right)\right\rangle =\hat{M}_{yz}\left|\Psi\left(z,p_{z}\right)\right\rangle $.
Being $\hat{M_{yz}}^{-1}\hat{L}_{x}\hat{M}_{yz}=-\hat{L}_{x}$, it
follows that $L_{x}\left(z,p_{y}\right)=-L_{x}\left(z,p_{z}\right)$.

\begin{figure}
	\centering
	\includegraphics[width=0.87\columnwidth]{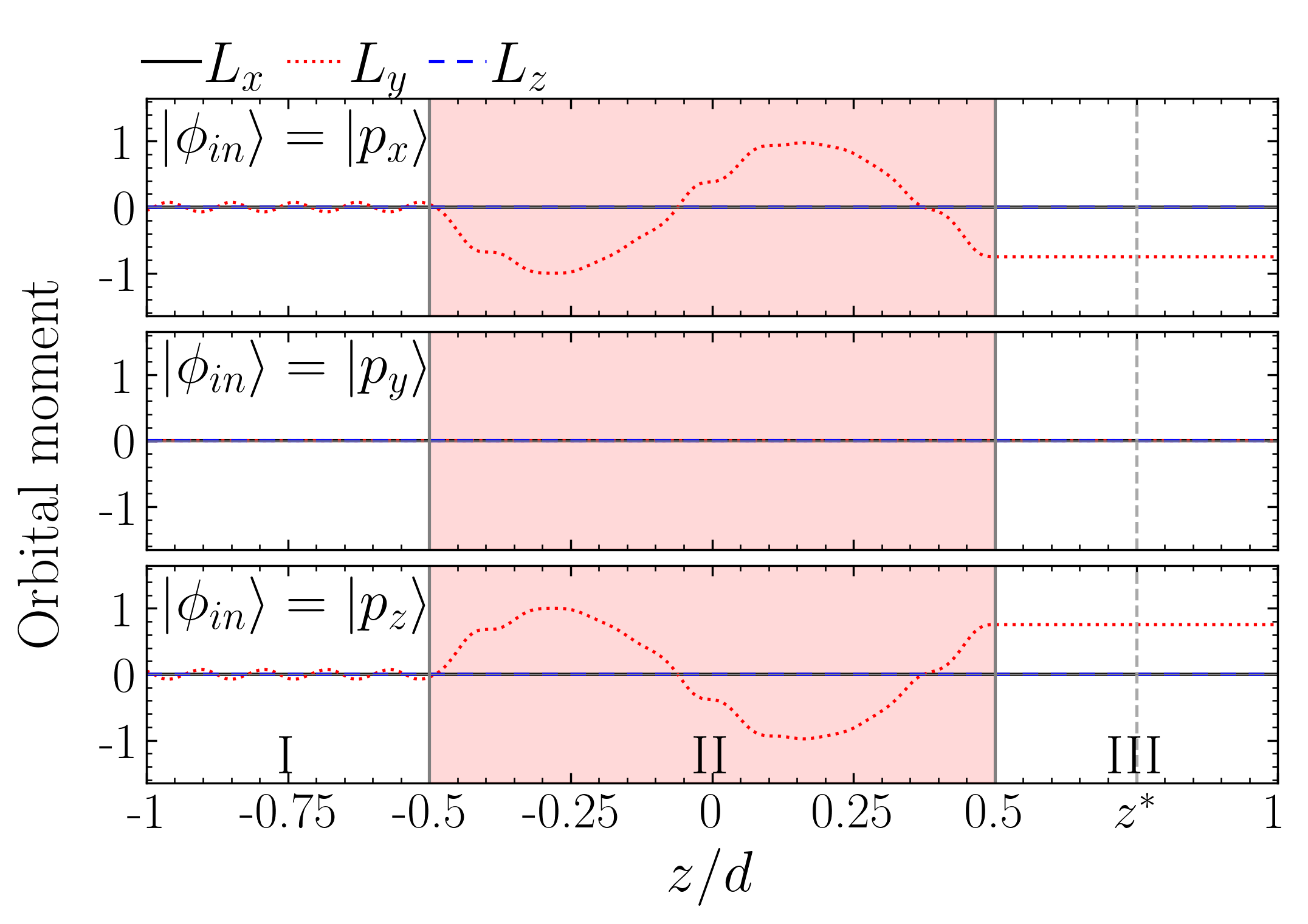}
	\caption{Spatial profile of $\boldsymbol{L}$ with \textcolor{black}{$\boldsymbol{g_{0}}/\epsilon_{0}=\left(0,0.5,0\right)$, $d=20$ nm, and $\mu/\epsilon_0=2$,} considering different initial configurations $\left|\phi_{in}\right\rangle$ ($\epsilon_0=38$ meV). As expected, when $\left|\phi_{in}\right\rangle =\left|p_{y}\right\rangle$, the orbital moment remains quenched. On the other hand, $L_{y}$ is the only nonvanishing component for the other initial configurations, where $L_{y}\left(z,p_{x}\right)=-L_{y}\left(z,p_{z}\right)$.}
	\label{fig:supp1}
\end{figure}

\begin{figure}
	\centering
	\includegraphics[width=0.87\columnwidth]{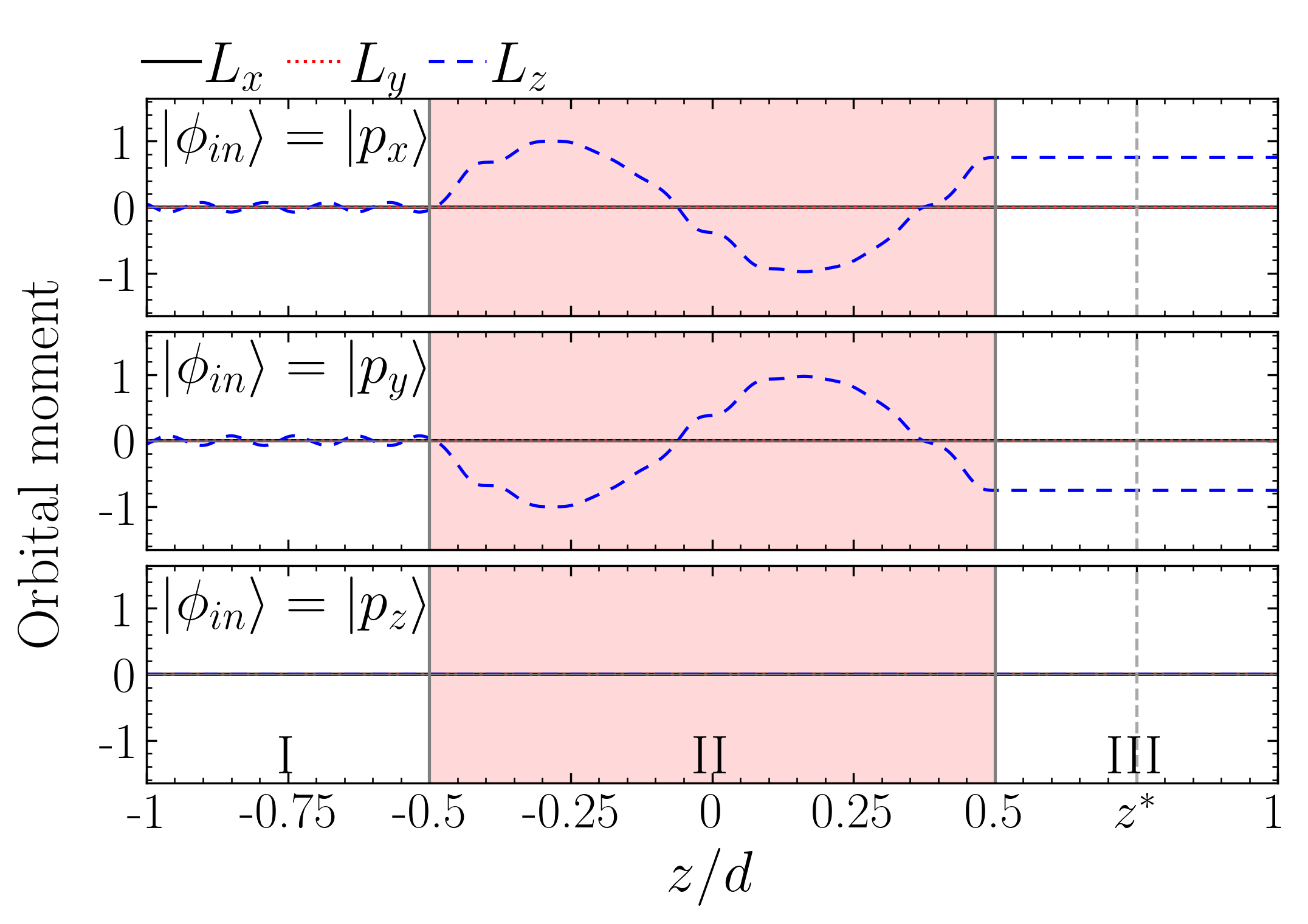}
	\caption{Spatial profile of $\boldsymbol{L}$ with \textcolor{black}{$\boldsymbol{g_{0}}/\epsilon_{0}=\left(0,0,0.5\right)$, $d=20$ nm, and $\mu/\epsilon_0=2$,} starting from three different initial configurations $\left|\phi_{in}\right\rangle$ ($\epsilon_0=38$ meV). If $\left|\phi_{in}\right\rangle =\left|p_{z}\right\rangle$, we do not observe any orbital polarization. On the contrary, $L_{z}$ is non-zero when $\left|\phi_{in}\right\rangle =\left|p_{x}\right\rangle, \left|p_{y}\right\rangle$. The preserved symmetries $\hat{M}_{xy}$ and $\hat{M}_{\overline{xy}}$ 
    dictate that $L_{z}\left(z,p_{x}\right)=-L_{z}\left(z,p_{y}\right)$.}
	\label{fig:supp2}
\end{figure}

We can reach analogous conclusions also when only \textcolor{black}{$g_{0y}\neq0$} or only \textcolor{black}{$g_{0z}\neq0$} (with the same parameters $a_{i}^{A}$ and $\Delta_{i}^{A}$). In the first case, the corresponding transmission matrix $\hat{\mathcal{T}}$ commutes with the symmetry operators $\hat{\mathcal{M}}_{y}$, $\hat{M}_{xz}$, and $\hat{M}_{\overline{xz}}$. It follows that the angular moment components are always vanishing in region $\text{III}$, except for $L_{y}$ when the electron is prepared in $\left|p_{x}\right\rangle$ or $\left|p_{z}\right\rangle$ (opposite values), as we can observe in Fig. \ref{fig:supp1}. On the other hand, when only $g_{0z}\neq0$, the preserved symmetries are $\hat{\mathcal{C}}_{2z}$, $\hat{M}_{xy}$, and $\hat{M}_{\overline{xy}}$. As a consequence, $L_{z}$ is the only nonvanishing component when $\left|\phi_{in}\right\rangle =\left|p_{x}\right\rangle ,\left|p_{y}\right\rangle$, with $L_{z}\left(z,p_{x}\right)=-L_{z}\left(z,p_{y}\right)$ (see Fig. \ref{fig:supp2}).

\begin{figure}
	\centering
	\includegraphics[width=0.9\columnwidth]{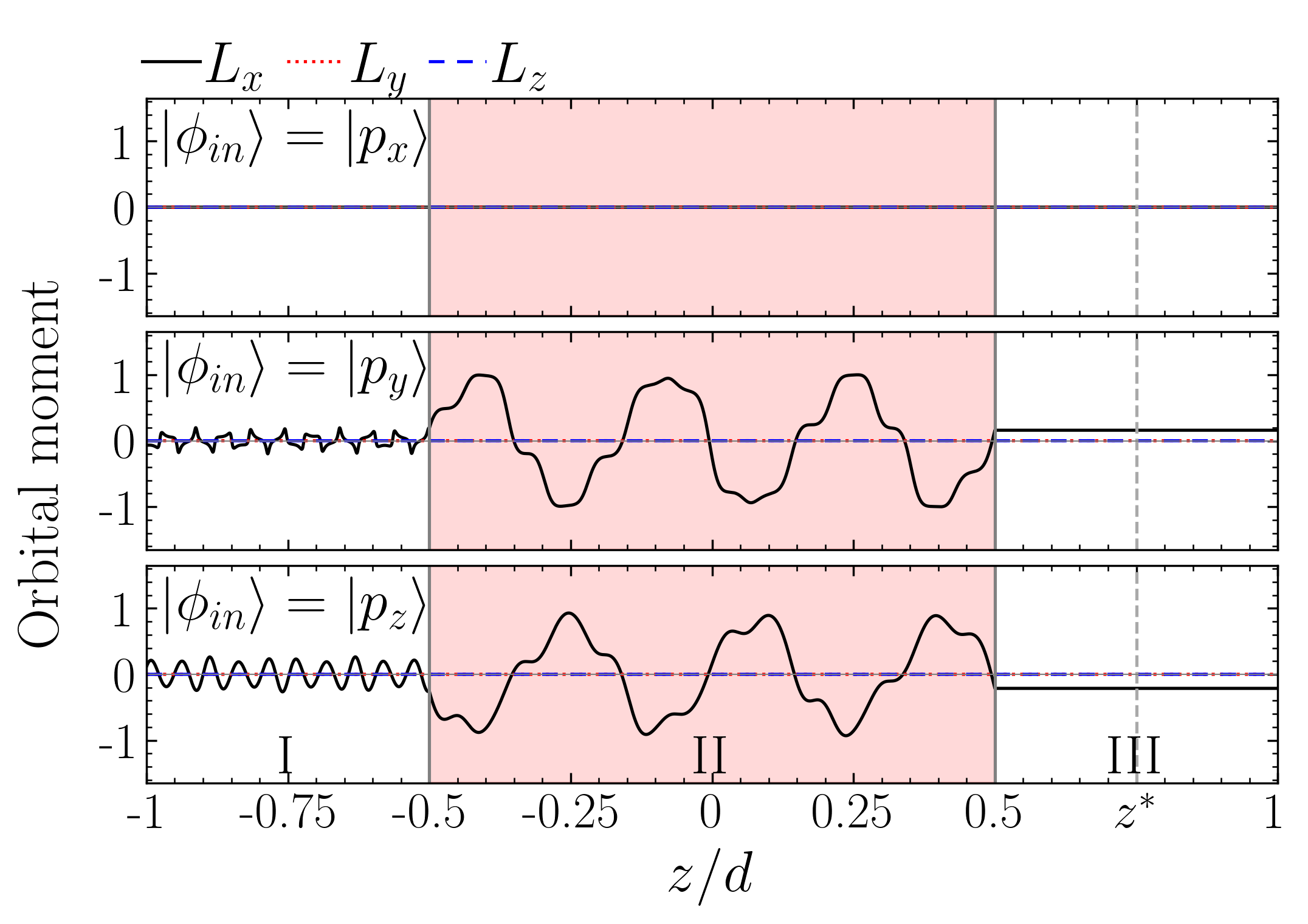}
	\caption{Spatial profile of $\boldsymbol{L}$ with \textcolor{black}{$\boldsymbol{g}/\epsilon_0=\left(0.5,0,0\right)$, $d=20$ nm, and $\mu/\epsilon_0=2$,} considering different initial configurations $\left|\phi_{in}\right\rangle$ ($\epsilon_0=38$ meV). While \textcolor{black}{$\boldsymbol{g_0}=\boldsymbol{0}$}, crystal-field effects are introduced in regions $\text{I}$ and $\text{II}$ through the following parameters: $\boldsymbol{a}^{\text{I}}=\left(0.3,0.6,-0.2\right)$, $\boldsymbol{\Delta}^{\text{I}}=\left(0.25,0.5,-0.7\right)$, $\boldsymbol{a}^{\text{II}}=\left(0.65,0.4,0.35\right)$, and $\boldsymbol{\Delta}^{\text{II}}=\left(-0.75,-0.25,0.8\right)$, where $\boldsymbol{a}^{A}=\left(a_{x}^{A},a_{y}^{A},a_{z}^{A}\right)$ and $\boldsymbol{\Delta}^{A}=\left(\Delta_{x}^{A},\Delta_{y}^{A},\Delta_{z}^{A}\right)$. Due to the preserved symmetry $\hat{\mathcal{M}}_x$, $L_{x}$ is the only nonvanishing component for $\left|\phi_{in}\right\rangle =\left|p_{y}\right\rangle ,\left|p_{z}\right\rangle $. Nevertheless, the symmetries $\hat{M}_{yz}$ and $\hat{M}_{\overline{yz}}$ are now broken  and, thus, $L_{x}\left(z,p_{y}\right)\neq-L_{x}\left(z,p_{z}\right)$.}
	\label{fig:supp3}
\end{figure}

Afterwards, let modify the Hamiltonian in Eq. \ref{eq:ham_gx_0} by introducing crystal-field effects in regions $\text{I}$ and $\text{II}$ through the parameters $a_{i}^{\text{I},\text{II}}$ and $\Delta_{i}^{\text{I},\text{II}}$, while preserving $\hat{\mathcal{M}}_{x}$. \textcolor{black}{In addition, we account for renormalization effects of the electron effective mass in region $\text{II}$ by switching-on $g_x$. In such a way, we have to require that $\hat{\mathcal{A}}^{\text{I}}D\left|\Psi^{\text{I}}\left(-d/2\right)\right\rangle =\hat{\mathcal{A}}^{\text{II}}D\left|\Psi^{\text{II}}\left(-d/2\right)\right\rangle$ and $\hat{\mathcal{A}}^{\text{II}}D\left|\Psi^{\text{II}}\left(d/2\right)\right\rangle =\hat{\mathcal{A}}^{\text{III}}D\left|\Psi^{\text{III}}\left(d/2\right)\right\rangle$.} As shown in Fig. \ref{fig:supp3}, the introduction of anisotropy through those parameters does not affect qualitatively the trend of the electronic angular moment in region $\text{III}$. Indeed, the only nonvanishing component is still $L_{x}$ if $\left|\phi_{in}\right\rangle =\left|p_{y}\right\rangle ,\left|p_{z}\right\rangle$, as dictated by $\hat{\mathcal{M}}_{x}$. On the contrary, $L_{x}\left(z,p_{y}\right)\neq-L_{x}\left(z,p_{z}\right)$ because we break both $\hat{M}_{yz}$ and $\hat{M}_{\overline{yz}}$ compared to the case discussed in the main text.

\begin{figure}
    \centering
\includegraphics[width=0.9\columnwidth]{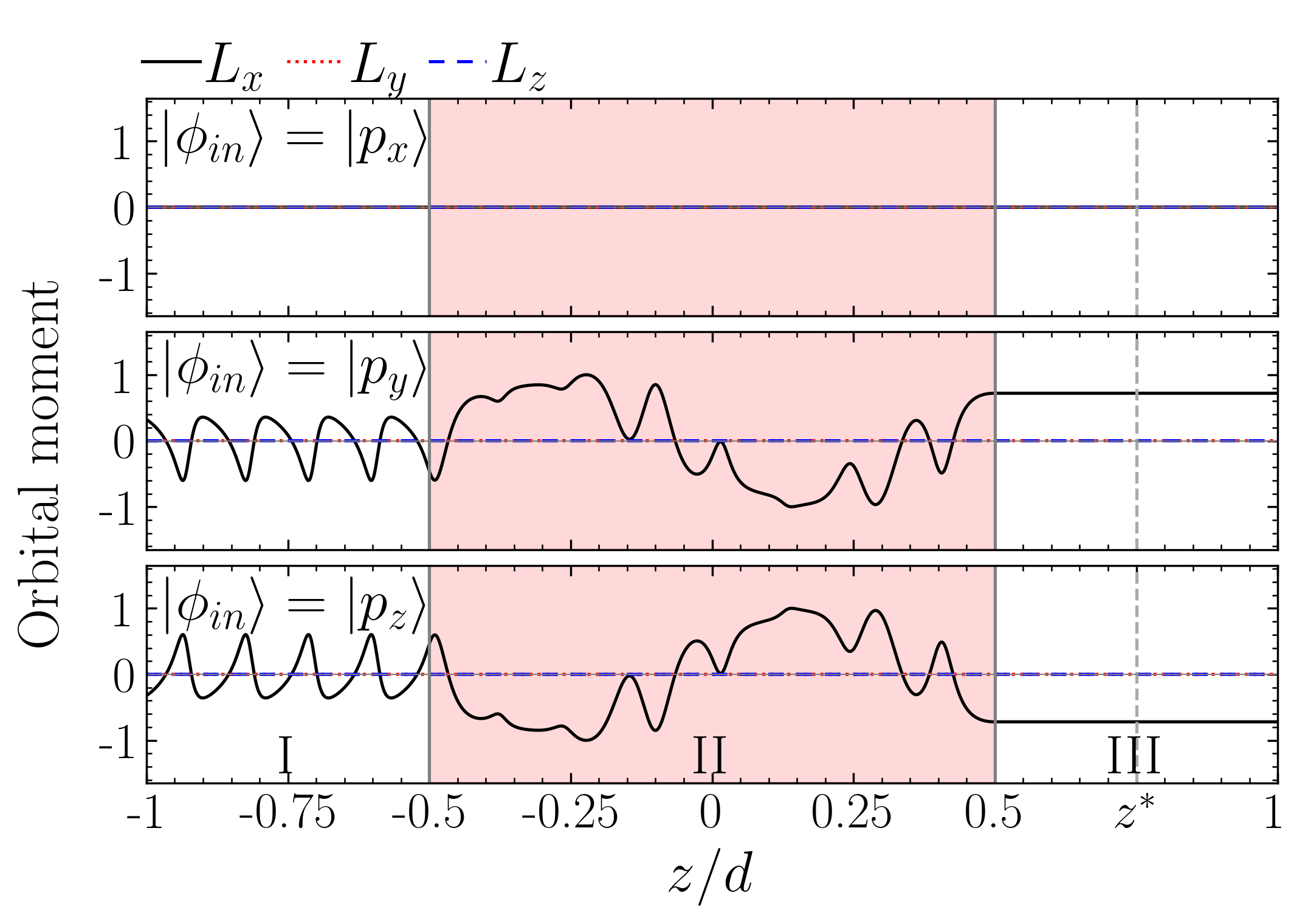}
    \textcolor{black}{\caption{Spatial resolved orbital moment $\boldsymbol{L}$ in the regions $\text{I}$, $\text{II}$ (pink shaded area), and $\text{III}$, considering different input configurations with $g_{0x}/\epsilon_0=0.5$ as the only non-vanishing mirror-breaking parameter ($\epsilon_0=38$ meV, $d=20$ nm, and $\mu/\epsilon_0=2$). Here, a delta function potential, $V(z)=\delta\left(z+d/2\right)+\delta\left(z-d/2\right)$, is considered at the interfaces, i.e., at $z=\pm d/2$.}
\label{fig:delta_interfaces}}
\end{figure}

\begin{figure}
\centering
\includegraphics[width=0.9\columnwidth]{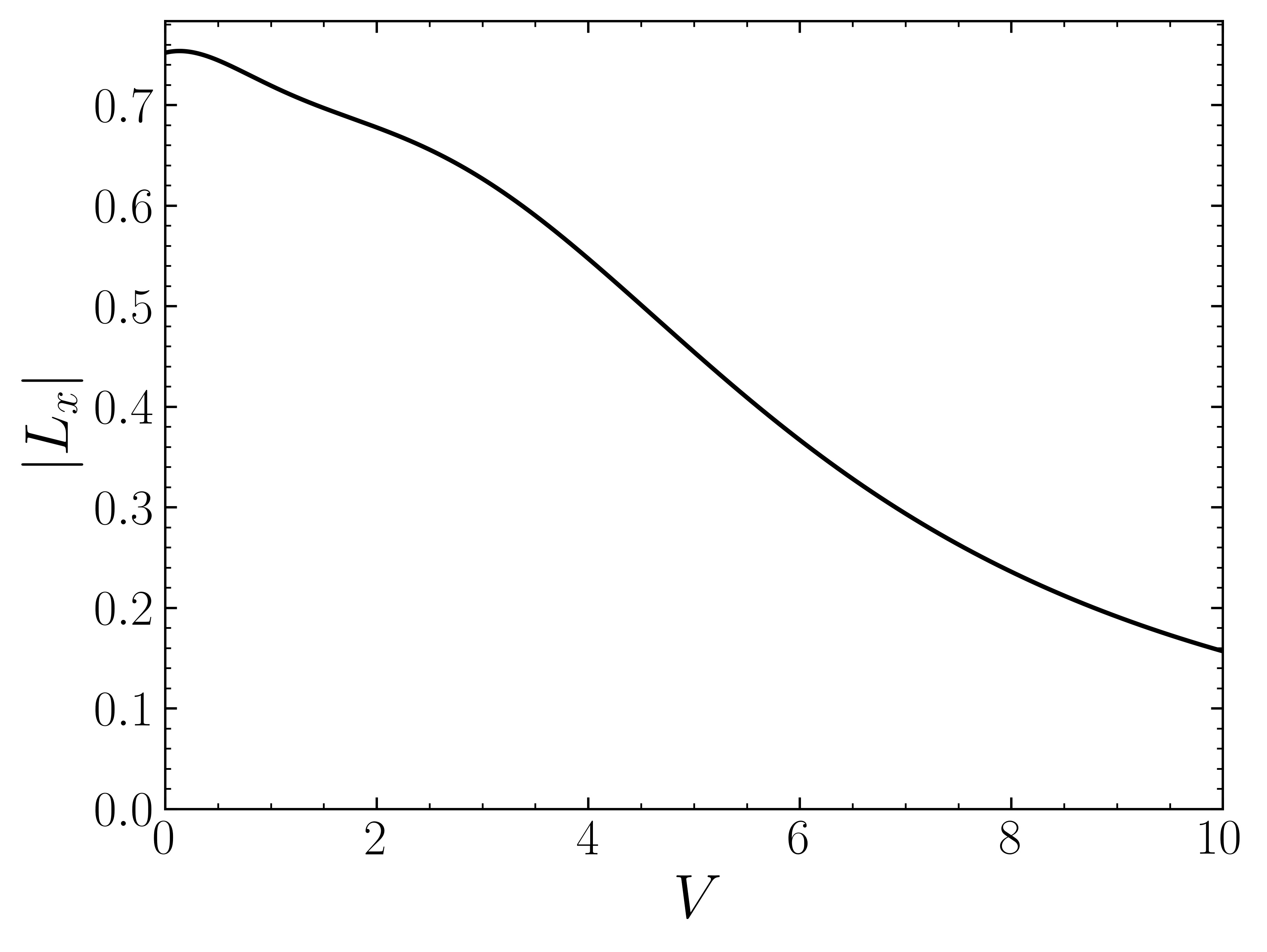}
    \textcolor{black}{\caption{Trend of the transmitted $|L_{x}|$ for different values of $V$, with $g_{0x}/\epsilon_0=0.5$ and $\phi_{in}=p_{z}$ ($\epsilon_0=38$ meV, $d=20$ nm, and $\mu/\epsilon_0=2$). Here, $V$ (in units of $\epsilon_{0}d$) represents the amplitude of the delta function potential at the interfaces: $V(z)=V[\delta\left(z+d/2\right)+\delta\left(z-d/2\right)]$. The amplitude of the filtered component decays as $V$ increases due to the suppression of the transmission amplitude.}
    \label{fig:trend_V}}
\end{figure}

\begin{figure}
    \centering
\includegraphics[width=0.9\columnwidth]{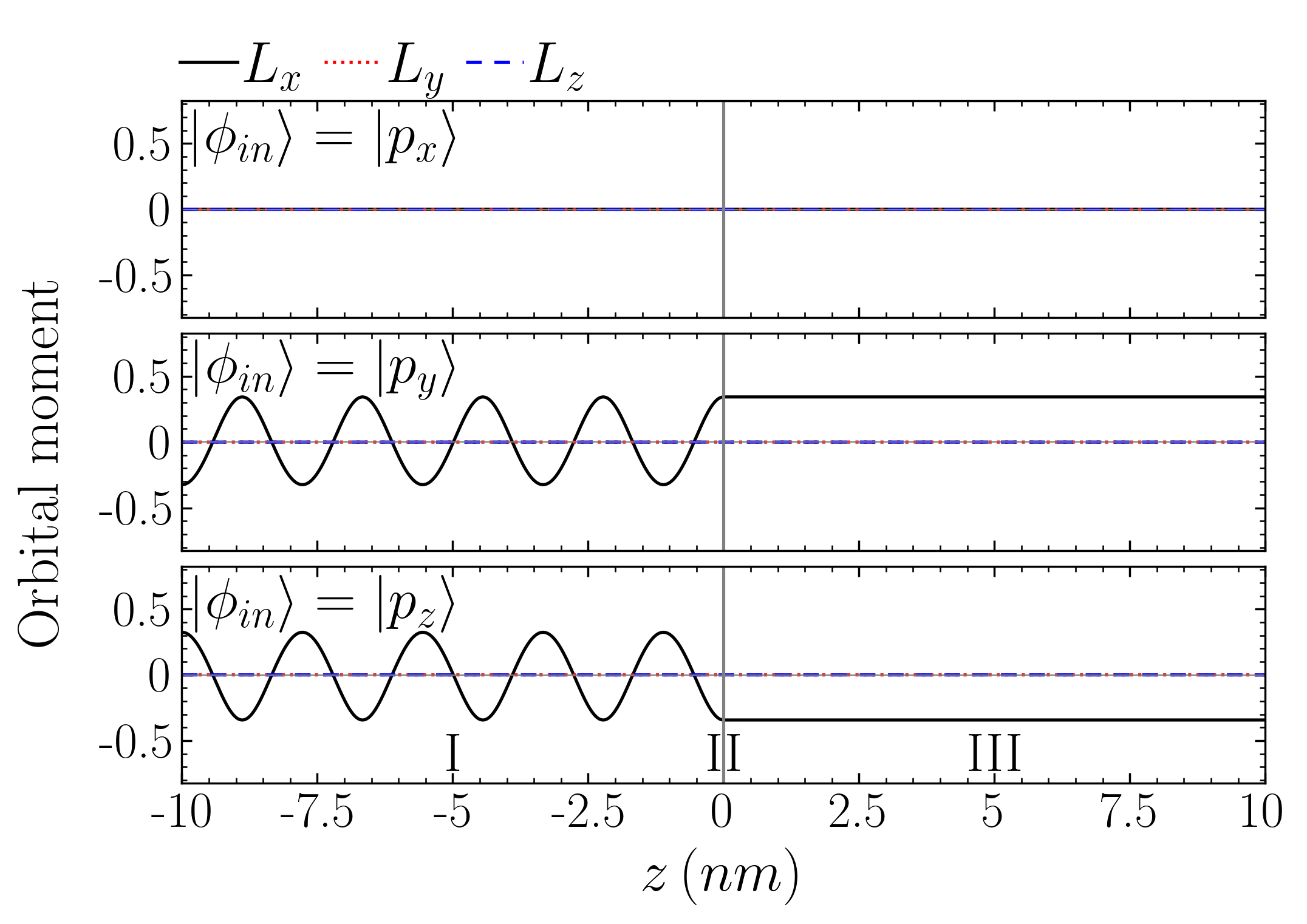}
    \textcolor{black}{\caption{Spatial profile of the  orbital moment $\boldsymbol{L}$ for different orbital configurations of the injected electrons with $\mu/\epsilon_0=2$ ($\epsilon_0=38$ meV). In this case, the region $\text{II}$ corresponds to a local orbital dependent scattering potential at $z=0$ with a delta function form, i.e., $V\left(z\right)=\delta\left(z\right)\hat{\chi}_{x}/2$.} 
\label{fig:delta_molecule}}
\end{figure}

\textcolor{black}{Next, let focus on the electronic transmission across a heterostructure described by the following Hamiltonian:
\begin{equation}
\hat{\mathcal{H}}\left(z\right)=\begin{cases}
\hat{p}_{z}^{2}\mathcal{I} & z<-d/2\\
\hat{p}_{z}^{2}\mathcal{I}+g_{0x}\hat{\chi}_{x}+V\left(z\right)\mathcal{I} & -d/2\leq z\leq d/2\\
\hat{p}_{z}^{2}\mathcal{I} & z>d/2
\end{cases},
\end{equation}
where a delta function  potential is considered at the interfaces: $V\left(z\right)=V_{1}\delta\left(z+d/2\right)+V_{2}\left(z-d/2\right)$, with $V_{1},V_{2}>0$ expressed in units of $\epsilon_0d$. In such a way, the derivatives of the wavefunctions are discontinuous at the interfaces and must satisfy the following conditions:
\begin{equation}
\begin{cases}
D\left|\Psi^{\text{I}}\left(-d/2\right)\right\rangle =D\left|\Psi^{\text{II}}\left(-d/2\right)\right\rangle -V_{1}\left|\Psi^{\text{II}}\left(-d/2\right)\right\rangle \\
D\left|\Psi^{\text{II}}\left(-d/2\right)\right\rangle =D\left|\Psi^{\text{III}}\left(-d/2\right)\right\rangle -V_{2}\left|\Psi^{\text{III}}\left(-d/2\right)\right\rangle 
\end{cases}.
\end{equation}
In any case, the introduction of a scattering delta potential at the interfaces does not alter the qualitative results regarding orbital filtering, as shown in Fig. \ref{fig:delta_interfaces}. Additionally, the behavior of the filtered $\left|L_x\right|$ is examined in terms of $V = V_1 = V_2$ in Fig. \ref{fig:trend_V}, revealing that, in the representative case under study, the amplitude of the transmitted component decays as high values of $V$ suppress the transmission amplitude. In conclusion, we investigate the case with regions $\text{I}$ and $\text{III}$ connected by an interface featuring an orbital-dependent
potential such that $V\left(z\right)=\tilde{g_0}\delta\left(z\right)\hat{\chi}_x$. Consequently, region $\text{II}$ effectively reduces to a point along the $z$-axis ($z=0$), where an orbital-dependent scattering potential is present. To solve the scattering problem, we have to impose that $\left|\Psi^{\text{I}}\left(0\right)\right\rangle =\left|\Psi^{\text{III}}\left(0\right)\right\rangle $ and $D\left|\Psi^{\text{I}}\left(0\right)\right\rangle=D\left|\Psi^{\text{III}}\left(0\right)\right\rangle -\tilde{g_{0}}\hat{\chi}_{x}\left|\Psi^{\text{III}}\left(0\right)\right\rangle $. In this case as well, the symmetry-breaking induced by local orbital couplings located at the interface leads to the filtering of the component $L_x$, as illustrated in Fig.~\ref{fig:delta_molecule}.}

\section{Nodal lines in the parameter space}\label{app:D}
Let consider the following Hamiltonian
\textcolor{black}{\begin{equation}
\hat{\mathcal{H}}\left(z\right)=
\begin{cases}
\hat{p}_{z}^{2}\text{\ensuremath{\mathcal{I}}} & z<-d/2\\
\hat{p}_{z}^{2}\text{\ensuremath{\mathcal{I}}}+g_{0x}\hat{\chi}_{x}+g_0\left(\hat{\chi}_{y}+\hat{\chi}_{z}\right) & -d/2\leq z\leq d/2\\
\hat{p}_{z}^{2}\text{\ensuremath{\mathcal{I}}} & z>d/2
\end{cases},
\end{equation}}
\textcolor{black}{with $g_{0x}\neq0$ and $g_0\neq0$.} In this case, it emerges clearly that all the mirror $\hat{\mathcal{M}}_{i}$ and rotation $\hat{\mathcal{C}}_{2i}$ transformations are broken in region $\text{II}$, while the symmetry $\hat{M}_{\overline{yz}}$ in the orbital space is preserved everywhere. Thus, if the initial orbital configuration is $\left|p_{x}\right\rangle$,
we infer that $\boldsymbol{g_0}\cdot\boldsymbol{L}\left(z,p_{x}\right)=0$.
Indeed, being $\hat{M}_{\overline{yz}}e^{ikz}\left|p_{x}\right\rangle=e^{ikz}\left|p_{x}\right\rangle$
and $\hat{M}_{\overline{yz}}^{-1}\hat{L}_{x}\hat{M}_{\overline{yz}}=-\hat{L}_{x}$,
we have that $L_{x}\left(z,p_{x}\right)=0$. Conversely,
because $\hat{M}_{\overline{yz}}^{-1}\hat{L}_{y}\hat{M}_{\overline{yz}}=-\hat{L}_{z}$,
we have that $L_{y}\left(z,p_{x}\right)=-L_{z}\left(z,p_{x}\right)$.
Therefore, it becomes evident that
\begin{equation}
\boldsymbol{g_0}\cdot\boldsymbol{L}\left(z,p_{x}\right)=g_0\left[L_{y}\left(z,p_{x}\right)+L_{z}\left(z,p_{x}\right)\right]=0.
\end{equation}
In a similar manner, by considering the operator $\hat{M}_{yz}$,
we can assert that $\boldsymbol{g_0}\cdot\boldsymbol{L}\left(z,p_{x}\right)=0$
if $g_{0y}=-g_{0z}$.
It means that, by providing a polar representation $\left(g_0,\theta_{\boldsymbol{g_0}},\phi_{\boldsymbol{g_0}}\right)$
of the vector $\boldsymbol{g_0}$ ($g_0$ as its modulus, $\theta_{\boldsymbol{g_0}}$
as its polar angle, and $\phi_{\boldsymbol{g_0}}$ as its azimuthal
one), we can clearly identify `robust' nodal lines in the parameter
plane $\left(\theta_{\boldsymbol{g_0}},\phi_{\boldsymbol{g_0}}\right)$
along which $\boldsymbol{g_0}$ and $\boldsymbol{L}\left(z,p_{x}\right)$
are orthogonal, regardless of the value of the modulus $g_0$. In fact,
with such a representation, we can impose $g_{0y}=\pm g_{0z}$ and get
two nodal lines dictated by $\cot\theta_{\boldsymbol{g_0}}=\pm\sin\phi_{\boldsymbol{g_0}}$.
This approach can be implemented also if the electron is prepared
in $\left|p_{z}\right\rangle$ ($\left|p_{y}\right\rangle$)
and $g_{0x}=\pm g_{0y}$ ($g_{0x}=\pm g_{0z}$), by considering $\hat{M}_{\overline{xy}}$ and $\hat{M}_{xy}$ ($\hat{M}_{\overline{xz}}$
and $\hat{M}_{xz}$), respectively. In particular, if the initial
orbital configuration is $\left|p_{z}\right\rangle$, straight
nodal lines appear in the parameter plane for $\phi_{\boldsymbol{g_0}}=\left(2k+1\right)\pi/4$
with $k=0,1,2,3$.

\begin{figure}[H]
	\centering
\includegraphics[width=0.9\columnwidth]{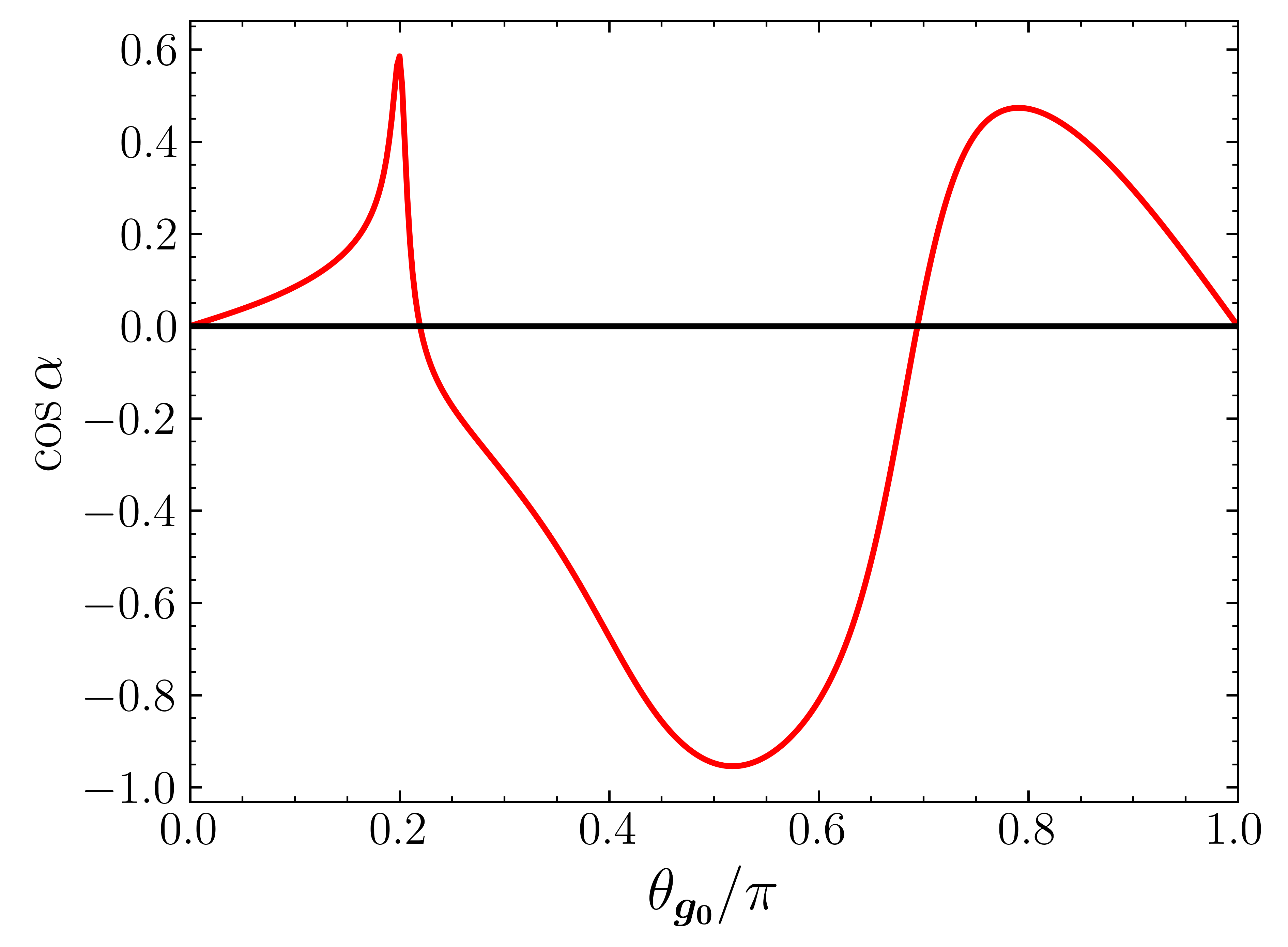}
	\caption{\textcolor{black}{Cosine of the angle $\alpha$ between $\boldsymbol{g_0}$ and $\boldsymbol{L}$ for the case where $\left|\phi_{in}\right\rangle = \left|p_z\right\rangle$, with $|\boldsymbol{g_0}|/\epsilon_0=0.5$ and $\phi_{\boldsymbol{g_0}} = \pi/4$, while varying $\theta_{\boldsymbol{g_0}}$ ($\epsilon_0=38$ meV and $d=20$ nm).} Two cases are considered: one (in black) where the symmetry $\hat{M}_{\overline{xz}}$ is preserved, and another (in red) where it is broken by $\boldsymbol{\Delta}^{\text{II}} = (-0.75, -0.25, 0.8)$. In the first case, $\boldsymbol{g_0} \perp \boldsymbol{L}$ ($\phi_{\boldsymbol{g_0}} = \pi/4$ defines a nodal line in the parameter space), while in the second one, the angle between $\boldsymbol{g_0}$ and $\boldsymbol{L}$ varies with $\theta_{\boldsymbol{g_0}}$ due to the absence of the $\hat{M}_{\overline{xz}}$ symmetry.
}
	\label{fig:supp4}
\end{figure}

As depicted in Fig. \ref{fig:supp4}, such nodal lines in the parameter space can be `weakened' by introducing strong anisotropy effects in the system through the terms proportional to $\hat{L}_{i}^{2}$ so as to break the involved symmetries $\hat{M}_{ij}$ and $\hat{M}_{\overline{ij}}$ (for example, by tuning the parameters $\Delta_{i}^{\text{II}}$).

%


\end{document}